\documentclass{aa}  

\usepackage{graphicx}
\usepackage{txfonts}
\usepackage{lipsum}
\usepackage{subcaption}         
\usepackage{lscape}             
\usepackage{placeins}           
\usepackage{tablefootnote}
\usepackage{array}

\usepackage[colorlinks=true,allcolors=blue]{hyperref}

\newcommand{\bs}{$\langle B \rangle$}
\newcommand{\vsini}{$v_\mathrm{e}\sin i$}
\newcommand{\kms}{km\,s$^{-1}$}
\newcommand{\orcidlink}[1]{\protect\href{https://orcid.org/#1}{\protect\includegraphics[width=8pt]{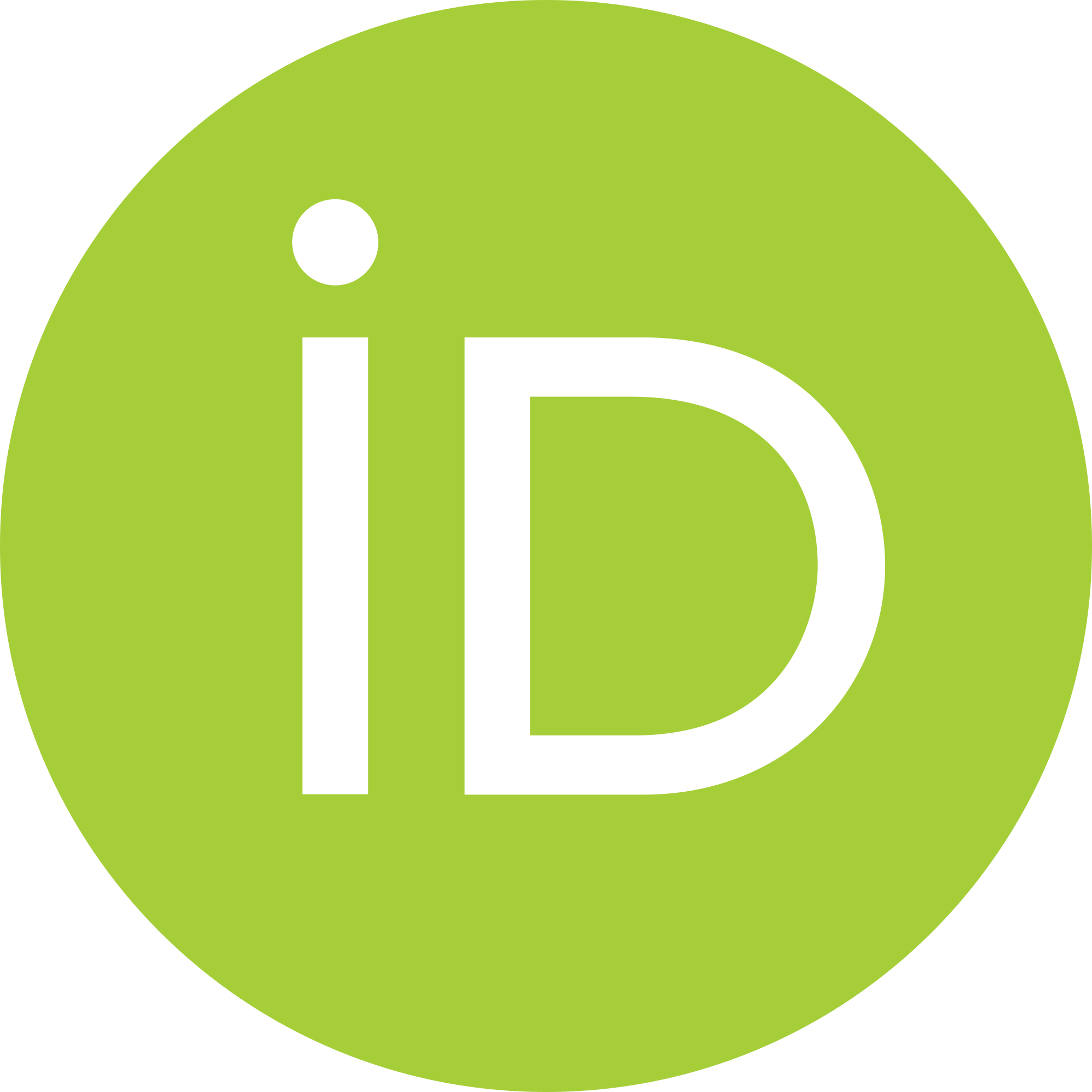}}}
\newcolumntype{C}{>{\centering\arraybackslash}m{0.5cm}}

\begin{document}

   \title{Testing the reliability of magnetic field strength \\measurements for M dwarfs}

   \author{I. Amateis\thanks{Corresponding author: \email{irene.amateis@physics.uu.se}}\inst{1}\orcidlink{0009-0007-0410-6215}
        \and O. Kochukhov\inst{1}\orcidlink{0000-0003-3061-4591}
        \and A. Hahlin\inst{2,1}\orcidlink{0000-0001-8365-8606} 
        }
        
   \authorrunning{Amateis, I., et al.} 
   \titlerunning{Testing the reliability of M-dwarf magnetic field measurements}

   \institute{Department of Physics and Astronomy, Uppsala University, Box 516, SE-751 20 Uppsala, Sweden
   \and Astrophysics Group, School of Chemical \& Physical Sciences, Keele University, Keele, Staffordshire ST5 5BG, UK}  

   \date{Received: 20 November 2025, Accepted: May 12, 2026}

\abstract
   {M dwarfs are known to host strong magnetic fields, which can be measured through several complementary techniques. However, the impact of key methodological choices on Zeeman broadening diagnostics has not been systematically quantified.}
   {Here we aim to assess the reliability of different approaches for inferring magnetic fields of M dwarfs and to identify strategies that yield the most accurate magnetic measurements.}
   {We combined state-of-the-art three-dimensional magnetohydrodynamic simulations of fully convective M dwarfs with MARCS model atmospheres to generate synthetic Stokes~$I$ spectra of a set of \ion{Ti}{i} lines. Synthetic observations were produced for different surface magnetic field strengths, projected rotational velocities, and inclination angles. Zeeman broadening and intensification were analysed using polarised radiative transfer calculations coupled with Markov chain Monte Carlo inference. We evaluated several statistical criteria (BIC, AIC, and WAIC) to determine the number of magnetic filling factors and compared two alternative strategies for treating line strengths.}
   {The inferred surface-averaged magnetic field is sensitive to the number of magnetic components. In most cases, BIC, AIC, and WAIC favoured the same number of components. In a few tests corresponding to more active and faster rotating stars, BIC and AIC favoured models with fewer components producing lower field estimates, while WAIC selected more complex models, which, generally, yielded a closer agreement with the input field strengths.
   Treating the intensity of each spectral line as a free parameter in the fitting process resulted in an underestimation of the field strength by 30–50\%, while fitting a joint element abundance along with continuum scaling recovered the input field more reliably. Moreover, we showed that the latter inference methodology adequately recovers the binned strength distribution of the magnetic field on the visible hemisphere.}
   {Zeeman broadening diagnostics can robustly recover magnetic fields in M dwarfs, but their accuracy strongly depends on methodological choices. Using quantitative statistical criteria for model selection and fitting continuum scaling factors together with element abundance provides reliable results and should be preferred in applications to observational data.}

\keywords{dynamo -- magnetic fields -- stars: activity -- stars: low-mass -- stars: magnetic field}

\maketitle
\nolinenumbers

\section{Introduction}
M dwarfs are characterised by their intense magnetic activity, which plays a crucial role in shaping both their internal structure and atmospheric properties. Unlike solar-like stars, which are thought to generate magnetic fields by dynamo mechanisms at the tachocline -- the boundary separating the radiative core and the convection zone \citep{Charbonneau2014} -- M dwarfs with masses below $\sim$\,0.35~$M_\odot$ have a fully convective interior structure, therefore lacking this transition region \citep{ChabrierBaraffe1997}. Despite early expectations that fully convective stars would produce only weak, small-scale magnetic fields \citep{Durney1993}, observations have revealed that M dwarfs commonly exhibit strong magnetic activity across a wide range of spatial scales \citep[for a detailed review see][]{Kochukhov2021}.

An accurate characterisation of magnetic fields in M dwarfs is essential to understand their physical properties, including the persistent and significant radius inflation observed in many systems \citep[see e.g.][]{Ribas2006}, which appears to correlate with the magnetic activity indicators \citep{LopezMorales2007} and has been theoretically explained by the magnetic inhibition of convection in stellar interiors \citep{Feiden2014}.

Another important area that requires detailed characterisation of M-dwarf magnetic fields is exoplanet research. Owing to their lower masses, planets orbiting these stars induce comparatively larger radial velocity signals than around more massive stars. Moreover, habitable-zone planets around M dwarfs orbit at shorter periods and with higher radial velocity amplitudes, facilitating their detectability. This makes M dwarfs prime targets for many ongoing and planned exoplanet studies. However, their magnetic activity must be taken into account and, ideally, mitigated to enable reliable detection of exoplanets using the radial velocity method \citep[e.g.][]{GomesdaSilva2012,Ruh2024} and to ensure accurate atmospheric characterisation through transit spectroscopy \citep[e.g.][]{Seager2024}. In addition, close-in habitable planets orbiting M-dwarf stars lie within the extended magnetospheres of their hosts, exposing them to the direct influence of stellar magnetic fields, magnetically driven stellar winds, and coronal mass ejections \citep{Vidotto2013,Vidotto2014,Kislyakova2017,AlvaradoGomez2022}. These interactions significantly affects both the atmospheres and internal structures of such exoplanets.

Several complementary methods have been developed to probe stellar magnetism. A commonly used approach is Zeeman Doppler imaging \citep[ZDI,][]{DonatiLandstreet2009}, which uses time-series polarimetric (Stokes $V$) spectra to reconstruct the large-scale magnetic field geometry. Alternatively, the absolute strength of the magnetic field can be inferred from Zeeman broadening and intensification in intensity (Stokes $I$) spectra \citep[see e.g.][]{Reiners2012}. Both effects originate from the magnetic splitting of spectral lines and the associated redistribution of line opacity among Zeeman components, and they represent two observational manifestations of the same physical process. The magnitude of the spectral line splitting depends on the magnetic field strength, the wavelength, and the effective Land\'e factor ($g_{\rm eff}$) of the transition. When rotational broadening is small, the splitting can be directly resolved, and the magnetic field is primarily inferred from Zeeman broadening of spectral lines. For more rapidly rotating stars, rotational broadening masks the Zeeman line splitting. As a result, the redistribution of opacity among Zeeman components constitutes the only observable magnetic signature, manifesting as an increase in the equivalent width of magnetically sensitive lines, commonly referred to as Zeeman intensification. 

The magnetic field inferred from broadening or intensification contains all contributions from the visible stellar surface, including from very small-scale magnetic components. This is in contrast with the ZDI technique which is sensitive to polarised light, and, as a consequence, contributions coming from surface elements with opposite magnetic polarity cancel each other out. For a more detailed description of the Zeeman broadening magnetic field inference methodology, we refer to \citet{Kochukhov2021}.

Studies comparing magnetic diagnostic results obtained with ZDI and Zeeman broadening for a wide range of stars have shown that magnetic fields measured from intensity spectra are always much stronger than those inferred from polarimetry, as the majority of the magnetic energy is carried within small-scale fields \citep{See2019,Lavail2019,Kochukhov2020,Kochukhov2021}. Therefore, a comprehensive understanding of stellar magnetic fields requires multiple diagnostic methods. However, when assessing the total magnetic field strength and the energetically dominant surface component, emphasis should be placed on the Zeeman broadening diagnostic.

The combination of strong surface magnetic fields in M dwarfs -- significantly exceeding those in FGK stars \citep{Reiners2012} -- the advent of high-resolution near-infrared spectrographs, where Zeeman splitting is enhanced due to its $\lambda^2$ wavelength dependence, and advancements in spectral synthesis and inference techniques have collectively driven major progress in M-dwarf magnetic field measurements from Stokes $I$ spectra. Over the past decade, the state of the art has advanced from measuring a single field strength or applying simple two-component models to employing sophisticated multi-line reconstructions of the magnetic field strength distribution, often incorporating numerous magnetic components \citep{Shulyak2017,Reiners2022,Cristofari2023}. These methodologies have also been extended to fast-rotating stars \citep{Kochukhov2017,Kochukhov2019,Shulyak2019}, where magnetic intensification becomes the dominant diagnostic over Zeeman broadening. Despite this progress, the accuracy, biases, and limitations of such Stokes $I$-based magnetic field studies of M dwarfs remain largely unquantified. Additionally, key methodological choices, such as magnetic field parametrisation and the treatment of non-magnetic parameters, have yet to be systematically explored. The aim of this work is to address these gaps by providing an assessment of the most commonly used Zeeman broadening techniques for magnetic field measurement based on a realistic magnetic field geometry appropriate for fully convective M dwarfs.

Significant efforts have been expended to create magnetohydrodynamic models aimed at deriving the full spatial power spectrum of magnetic field in fully convective M dwarfs. In particular, \citet{Yadav2015} published a high-resolution three-dimensional numerical simulation of a fully convective low-mass star. This simulation provided a realistic magnetic field topology that included both a large-scale dipole-like axisymmetric structure similar to global fields commonly found in M dwarfs and small-scale distributed fields. Here we use the corresponding surface magnetic field map to generate simulated observations across a range of magnetic field strengths and rotational parameters, and then test the recovery of magnetic characteristics from intensity spectra using multi-component magnetic filling factor models. Specifically, we explore the impact of different assumptions about the number of magnetic components on the inferred field strengths.

This paper is organised as follows. Section \ref{sub_sec:MHDmodel} describes in more detail the theoretical magnetic field geometry model adopted from the study by \citet{Yadav2015}, while Sect.~\ref{sub_sec:SimulatedObs} explains how this MHD model topology was used to produce simulated observations. Section~\ref{sub_sec:Inference} presents the method of extracting information on the magnetic field. Results and their analysis are discussed in Sect.~\ref{Sec:Results}. The general conclusions of our study are summarised in Sect.~\ref{sec:Conclusions}.

\section{Methods}
\label{sec:Methods}

\subsection{MHD model}
\label{sub_sec:MHDmodel}

In this work, we employed a realistic, physically motivated model of the magnetic field in a low-mass fully convective star to test Zeeman broadening magnetic field inference techniques. To this end, we adopted a surface field topology corresponding to a snapshot of the theoretical 3D magnetohydrodynamic (MHD) calculations published by \citet{Yadav2015}. These authors conducted simulations of an $\alpha^2$-dynamo based on the anelastic MHD equations in a highly stratified convective sphere with characteristics resembling those of a low-mass star. Notably, both the density contrast and the surface angular resolution of their calculations surpass those of other MHD models of fully convective stars \citep{Dobler2006,Browning2008,Kapyla2021,Ortiz-Rodriguez2023}. Moreover, \citet{Yadav2015} configured their simulations with the explicit aim of producing realistic magnetic field strengths, which is particularly valuable for our investigation. To date, their model remains the only one that has demonstrated the ability to simultaneously reproduce the typical magnetic field strengths of M dwarfs at both large and small spatial scales.

\begin{figure*}[t!]
\centering
\includegraphics[width=\hsize]{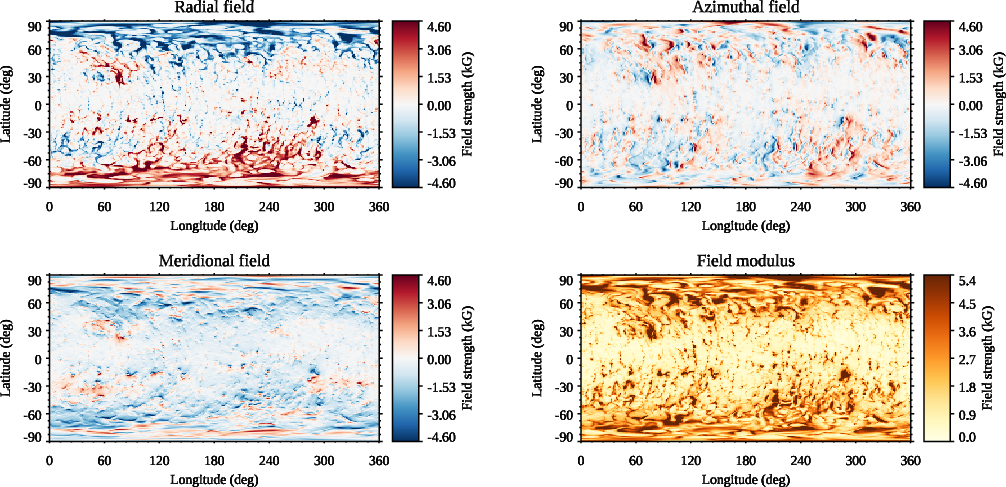}
\caption{MHD magnetic field structure from \citet{Yadav2015} employed in our study. The four panels display rectangular maps of the radial, meridional, and azimuthal field components as well as the field modulus. Colour bars span the range encompassing 95\% of the data values.}
\label{fig:MHD} 
\end{figure*}

The surface field topology resulting from the \citet{Yadav2015} model is illustrated in Fig.~\ref{fig:MHD}, which shows rectangular projections of the radial, meridional, and azimuthal field components, complemented by a map of the field modulus. The surface magnetic field strength spans approximately $\pm16$~kG. As seen in Fig.~\ref{fig:MHD}, the MHD model generates a strong, dipolar-like global magnetic field. \citet{Yadav2015} have already demonstrated, by applying ZDI to simulated Stokes $V$ observations, that inversions based on spectropolarimetric time series can successfully recover this global topology. At the same time, vigorous convection in the surface layers shreds this global field, producing an intricate pattern of small-scale magnetic structures that dominates the field strength distribution at smaller spatial scales. The aim of our study is to assess how accurately key characteristics of this small-scale magnetic field can be recovered.

The MHD model computed by \citet{Yadav2015} corresponds to a moderate stellar rotation with a period of 20~d, and with an average surface field strength of 1.584~kG. This is representative of an M dwarf near the threshold where stellar activity begins to saturate \citep{Reiners2022}.

\subsection{Simulated observations}
\label{sub_sec:SimulatedObs}

In this section we discuss how, relying on the MHD model described in Sect.~\ref{sub_sec:MHDmodel}, we   simulated observations. 

The selection of spectral lines employed for field strength measurements is an important aspect of our study. In this work, we used a set of spectral lines that \cite{Kochukhov2017} and \cite{Shulyak2017} showed to be particularly useful for Zeeman broadening and intensification analyses. This set is a multiplet of \ion{Ti}{i} lines corresponding to the transitions between the $\mathrm{^5F}$ and $\mathrm{^5F^o}$ energy levels, with wavelengths in the $\lambda$ 964.74--978.77 nm range (see Table~\ref{tab:ti_lines}). The magnetic sensitivity changes significantly from one line to the other due to different Land\'e factors and Zeeman splitting patterns. Additionally, one of these \ion{Ti}{i} lines, $\lambda$ 974.36 nm, has zero effective Land\'e factor, making this line useful to constrain titanium abundance and disentangle magnetic broadening from other broadening mechanisms, such as rotational broadening. Lastly, this set of \ion{Ti}{i} is relatively free from molecular blends, even in late-M dwarfs.

Despite its unique characteristics and high resolution, the MHD calculation by \citet{Yadav2015} remains unable to resolve the outermost stellar layers and lacks essential physical ingredients (such as a realistic equation of state, detailed opacities and non-grey radiative transfer) required to provide a realistic velocity field and temperature–density profile for spectrum synthesis. Consequently, we adopted a single plane-parallel \textsc{MARCS} \citep{Gustafsson2008} model atmosphere with an effective temperature of $T_{\rm eff}=3400$~K and a surface gravity of $\log g=5.0$ for spectral modelling. These atmospheric parameters correspond to the middle of the $T_{\rm eff}$ range explored in previous M-dwarf magnetic studies and are consistent with an M3.3V star at the fully convective boundary \citep{Pecaut2013}, in agreement with the interior structure assumed by \citet{Yadav2015}.

Using the same thermodynamic structure for surface regions with varying field strengths is undoubtedly a simplification. Nevertheless, both observational evidence \citep{Berdyugina2005} and local MHD simulations of M-dwarf atmospheres \citep{Beeck2015} indicate that temperature and brightness contrasts between magnetic and non-magnetic regions in these stars are considerably reduced compared to those in active G and K stars. No significant velocity fluctuations are expected in M-dwarf atmospheres according to local 3D HD models \citep{Wende2009}, leading us to adopt zero micro and macroturbulent velocities.

The \ion{Ti}{i} line parameters, along with information on minor atomic blends affecting these lines, were extracted from the VALD database \citep{Ryabchikova2015} using the aforementioned MARCS model atmosphere and the solar chemical abundance table from \citet{Asplund2021}. Molecular blends were not included in the calculations, as the primary molecular contributors, TiO and FeH, are both relatively weak at $T_{\rm eff}=3400$~K around the studied \ion{Ti}{i} multiplet, with TiO lacking accurate line lists in this wavelength region. This omission of molecular absorption is consistent with the treatment of these neutral titanium lines in observational studies \citep[e.g. see][]{Shulyak2017,Shulyak2019,Reiners2022}.

We began by computing synthetic Stokes spectra using \textsc{MARCS} model atmospheres \citep{Gustafsson2008} in combination with the polarised radiative transfer code \textsc{Synmast} \citep{Kochukhov2010}. These calculations were performed over a fine grid of magnetic field strengths, limb angles, and field vector orientations relative to the line of sight. For each surface element in the fully resolved MHD field structure, we then applied linear interpolation within this grid to derive the local spectrum corresponding to the field vector obtained from the MHD field map. Subsequently, disk-integrated spectra were computed by summing the contributions from all visible surface elements, accounting for their projected areas, and Doppler shifts due to solid-body rotation. These calculations were repeated for ten evenly spaced rotational phases, three inclination angles ($i=15, 45, 75$\degr), and projected rotational velocities (\vsini) ranging from 1 to 20~\kms. 

To simulate realistic observations, the synthetic spectra were convolved with the instrumental profile assuming a spectral resolution of $R=65\,000$, and resampled to a velocity bin size of 2.0~\kms. These parameters are consistent with the characteristics of ESPaDOnS and Narval -- two high-resolution spectropolarimeters widely used for M-dwarf magnetic field studies \citep{Donati2008,Morin2008,Morin2010}. Gaussian noise was then added to achieve a target signal-to-noise ratio (S/N) of 100. This combination of line selection, spectral resolution, and S/N represents a conservative configuration, reflective of the initial analyses employing these \ion{Ti}{i} lines \citep{Kochukhov2017,Shulyak2017}. More recent investigations, such as that by \citet{Reiners2022}, have utilised spectra with a somewhat higher resolving power and significantly improved S/N, achieved by averaging observed spectra from several rotation phases. In Sect.~\ref{sub_sec:results_comparisonR22}, we analysed spectra with this enhanced data quality.

Since we aim to test the capabilities of field strength inference methods across a range of rotation rates and field intensities, we generated simulated observations by uniformly scaling the surface magnetic field of the original MHD calculation. We refer to this scaling as the magnetic scaling factor, $s$. This procedure allowed us to consider stars with weaker or stronger magnetic activity, while assuming that variations in field intensity are not accompanied by significant changes in the field morphology over a limited range of scaling factors. In this work, we produced spectra for $s=0.5, 1, 2$, corresponding to half, equal, and double the original magnetic field strength.

\begin{table}[]
    \centering
    \caption{\ion{Ti}{i} lines used for Zeeman broadening analysis.}
    \begin{tabular}{l c c c c}
    \hline \hline
    $\lambda$ (\AA) & $E$ (eV) & $\log(gf)$ & $g_\mathrm{eff}$ & $N_\mathrm{z}$\\
    \hline
    9647.37 & 0.818 & $-1.434$ & 1.53 & 15 \\
    9675.54 & 0.836 & $-0.804$ & 1.35 & 24 \\
    9688.87 & 0.813 & $-1.610$ & 1.50 & 9 \\
    9705.66 & 0.826 & $-1.009$ & 1.26 & 18 \\
    9728.41 & 0.818 & $-1.306$ & 1.00 & 12 \\
    9743.61 & 0.813 & $-1.306$ & 0.00 & 3 \\
    9770.30 & 0.848 & $-1.581$ & 1.55 & 27 \\
    9783.31\tablefootmark{a} & 0.836 & $-1.428$ & 1.49 & 21 \\
    9783.59\tablefootmark{a} & 0.818 & $-1.617$ & 1.49 & 9 \\
    9787.69 & 0.826 & $-1.444$ & 1.50 & 15 \\
    \hline
    \end{tabular}
\tablefoot{The columns give the central wavelength of the transition $\lambda$, the energy of the lower level $E$, the oscillator strength $\log(gf)$, the effective Land\'e factor $g_\mathrm{eff}$, and the total number of Zeeman components $N_{\rm z}$.\\
\tablefoottext{a}{These two lines are blended together and were analysed as a single feature.}
}
    \label{tab:ti_lines}
\end{table}

\subsection{Magnetic field inference}
\label{sub_sec:Inference}

This section describes how we infer magnetic field strengths and stellar parameters from the simulated observations presented in Sect.~\ref{sub_sec:SimulatedObs}. To model magnetically perturbed stellar spectra, we interpolated within the same synthetic library of local intensity spectra described in Sect.~\ref{sub_sec:SimulatedObs}, coupling it with a simplified representation of the stellar surface magnetic field. This representation, widely used in modelling Stokes $I$ spectra of active stars and proven to be sufficient for that purpose \citep[e.g.][]{Johns-Krull2007,Kochukhov2007,Shulyak2014,Cristofari2023}, assumes a uniform, purely radial magnetic field across the stellar surface. This simplification reduces the number of required local profile calculations to a small set -- seven limb angles in our case -- while preserving sufficient accuracy of the disk-integrated spectra, even for rapidly rotating stars.

Following a commonly employed strategy, the magnetic field is parametrised with a set of $N$ filling factors. A magnetic filling factor, denoted by $f_i$, represents the fraction of the stellar surface covered with a magnetic field of strength $B_i$. Therefore, the total spectrum is given by a linear combination,
\begin{equation}
    \label{eq:filling_fac_spectrum}
    S_\mathrm{tot}=\sum^N_{i}f_i S(B_i),
\end{equation}
of the disk-integrated spectra $S(B_i)$ corresponding to a set of field strength values $B_i$. In this work, the field components are equally spaced, starting at 0~kG and ending at $B_{\rm max} = B_N$. 

To prevent the filling factors from covering more or less than the entire surface of the star, we normalise them by using the following constraint:
\begin{equation}
    \label{eq:filling_fac_norm} 
    \sum^N_{i}f_i=1 \, .
\end{equation}
Under this condition, the magnetic field strength distribution is described by $N-1$ free parameters, sampled at equal intervals. The choice of sampling step is primarily determined by the spectral resolution and the wavelength coverage of the observations. Based on previous studies, a 2~kG step is appropriate for the simulated observations considered here, which correspond to a resolving power of $R = 65\,000$, a moderate S/N and wavelengths shorter than 1~$\mu$m. In contrast, data with higher resolution ($R \approx 10^5$), higher S/N ($\ga 200$) and longer wavelengths (1--2.5~$\mu$m) may require finer sampling \citep[e.g. see][]{Lavail2019,Hahlin2023}.

To address uncertainties in continuum normalisation common in the analyses of heavily blended spectra of M dwarfs, we used independent multiplicative continuum scaling factors for each spectral line employed in our analysis (see Table \ref{tab:ti_lines}). These scaling factors adjust the local continuum level of the synthetic spectrum. Each scaling factor multiplies the model flux in its corresponding wavelength region, leaving the line profile shape intact while allowing a small correction to the continuum.

In fitting each spectral line, we selected a wavelength interval that encompassed the line itself along with a portion of the surrounding continuum. In most cases, a fixed interval of 2.4~\AA\ was used; however, for calculations with \vsini\,=\,20~\kms, we extended the interval to 2.8~\AA\ to accommodate the increased line broadening caused by stellar rotation. As evident in Table~\ref{tab:ti_lines}, the lines at $\lambda$~9783.31~\AA\ and $\lambda$~9783.59~\AA\ are closely spaced. Therefore, we defined a single fitting interval for both and applied a common continuum scaling factor.

For the estimation of the set of stellar and magnetic field parameters, we performed a Markov Chain Monte Carlo (MCMC) sampling using the SoBAT library for IDL \citep{Anfinogentov2021} modified by \citet{Hahlin2022,Hahlin2023}. This approach provides realistic confidence intervals of the parameters, accounting for potential correlations, and allows one to easily incorporate external Bayesian constraints on any of the parameters.

We inferred the entire set of $N-1$ independent magnetic filling factors for each of the ten simulated rotation phases individually together with auxiliary stellar parameters, namely the titanium abundance $\epsilon (Ti) = \log{(N_{\rm Ti}/N_{\rm tot})}$, the projected rotational velocity \vsini, and nine continuum scaling factors. All these parameters were treated as free variables with uniform priors in the MCMC inference to account for degeneracies. The Ti abundance was included as a global factor controlling the line depths, since abundance variations can mimic the effect of magnetic intensification. The projected rotational velocity affects the width and shape of spectral lines and was thus fitted simultaneously with the magnetic field to correctly disentangle the rotational and Zeeman broadening effects.

The MCMC sampling started with 5000 burn-in steps and then continued until 100 independent samples were reached \citep[see details in][]{Hahlin2022,Hahlin2023}. The total number of steps in the parameter space varied from 30\,000 to over 200\,000 when analysing more active simulated stars, ensuring sufficient sampling around the region of the maximum likelihood in the parameter space. The final parameter values were calculated as medians of posterior distributions. The corresponding uncertainties were estimated from a confidence interval encompassing 68\% of the posterior distributions. The two dependent parameters of interest, including one filling factor and the mean field strength,
\begin{equation}
\langle B \rangle = \sum_i f_i B_i,
\end{equation}
were determined from the full posterior distributions, retaining information on correlations and degeneracies.

The optimal number of magnetic filling factors and sampling step size are not known a priori. Moreover, these parameters may differ depending on the quality of observational data, the overall stellar magnetic activity level, the rotational broadening, and the actual (unknown) shape of the underlying magnetic field strength distribution. Previous studies used different approaches to select the filling factors.

For example, \citet{Shulyak2017} and \citet{Shulyak2019} used a fixed number of magnetic components (11 $B_i$ values distributed between 0 and 20~kG with a 2~kG step) for all stars, excluding those with large \vsini\, for which the magnetic field recovery was more complicated. In those cases, they assumed a simpler two-component model: non-magnetic and magnetic with a single value of filling factor. Additionally, in some cases \citet{Shulyak2019} manually modified their filling factor distributions to remove the strong-field components that they deemed to be unreliable. On the other hand, \citet{Reiners2022} distinguished between inactive or moderately active stars, and more active M dwarfs. For the first class, they used magnetic components in the 0--4~kG range. For more active stars, they included components up to 12~kG. Moreover, the field strength step size was chosen to be 1~kG for stars with \vsini\,$<$\,5~\kms\ but 2~kG for stars with a higher rotational broadening. The authors insisted that these variations in the field parametrisation did not significantly influence the results. Finally, \citet{Cristofari2023} analysed a sample of inactive M dwarfs assuming field components from 0 to 10~kG in steps of 2~kG independently of stellar characteristics.

In contrast to these ad hoc approaches, \citet{Lavail2019} and \citet{Hahlin2022} advocated for a quantitative and objective method for selecting the number of filling factors. They computed the Bayesian information criterion \citep[BIC,][]{Sharma2017},
\begin{equation}
    \label{eq:BIC}
    \mathrm{BIC} = -2 \ln p(Y|\hat{\theta}) + d \ln n,
\end{equation}
for inferences with different number of filling factors. In this equation, $\ln p(Y|\hat{\theta})$ is the likelihood of the best fit for the variable $Y$ using parameters $\hat{\theta}$, $d$ is the number of parameters used in the inference, and $n$ is the number of data points. By increasing the number of parameters, the fit is more accurate, and consequently, the likelihood increases. However, one would face the risk of over-fitting. Eq.~(\ref{eq:BIC}) is meant to balance these two aspects by selecting the model with the lowest BIC.

BIC is but one of several possible information criteria that can be employed for model selection. Other examples include the Akaike information criterion (AIC),
\begin{equation}
    \label{eq:AIC}
    \mathrm{AIC} = -2 \ln p(Y|\hat{\theta}) + 2 d,
\end{equation}
which, compared to BIC, is less conservative and favours more complex models. A fully Bayesian generalisation of AIC is the Watanabe-Akaike information criterion (WAIC),
\begin{equation}
    \label{eq:WAIC}
    \mathrm{WAIC} =  - 2 \sum^n_i \ln \mathrm{E}_\theta [p(y_i|\theta)] + 2 \sum^n_i \mathrm{Var}_\theta [\ln p(y_i|\theta)]  
\end{equation}
where $\mathrm{E}_\theta [y_i|\theta]$ denotes the Bayesian expectation over the posterior distribution and $\mathrm{Var}_\theta$ the corresponding variance. Unlike the previously mentioned criteria, WAIC estimates the effective degrees of freedom from the likelihood function of the data and samples of the parameter $\theta$ obtained from the posterior distribution. Additionally, while BIC and AIC use a point estimate for the parameter to compute predictive density, WAIC uses the Bayesian predictive density.

The parameter space of our simulations (e.g. stellar properties, S/N, spectral resolution, line selection, $B_\mathrm{max}$, step size) is too large to explore exhaustively. Instead, here we adopt setups that reflect common practice and typical data characteristics, with the goal of assessing the reliability of existing observational studies. Our baseline uses a uniform field grid with 2~kG steps, as widely applied in the literature, with the choice of $B_\mathrm{max}$ governed by objective statistical criteria, BIC, AIC, and WAIC. The alternative approach of \citet{Reiners2022} -- employing 1~kG steps and a distinct line-scaling procedure -- is considered separately due to its methodological differences and application to a large but parameter-restricted sample of low-activity M dwarfs.

\section{Results}
\label{Sec:Results}

In Section \ref{sub_sec:MHDmodel}, we presented the MHD model simulating a fully convective M dwarf with a full-surface average magnetic field of \bs$_s \approx 1.6$~kG. In this work, the field distribution resulting from MHD calculations was scaled up and down to produce simulated observations, as described in Sect.~\ref{sub_sec:SimulatedObs}. Table \ref{tab:input_par} summarises the cases analysed in our work. Note that for the $s=2$ case, we tested \vsini\, between 3 and 20~\kms, while for $s=0.5$ we only considered \vsini\,=\,1, 3, and 5~\kms\, in order to reflect the fact that highly active stars are generally rotating faster than inactive M dwarfs. The table includes, for each set of input parameters, the true hemispheric average field strength, averaged over rotational phases (\bs$_{p,\mathrm{input}}$). The hemispheric averages were derived from the fully resolved MHD map, taking into account the adopted stellar inclination, projected surface areas, and using a linear limb-darkening law with $\eta=0.35$ as a weight. Furthermore, as a summary of our results, which are discussed in more detail in this section, the sixth column of Table~\ref{tab:input_par} presents the inferred mean hemispheric field averaged over the rotation phases (\bs$_p$) and the corresponding average upper and lower uncertainties from the inference. The seventh column shows the rotation-phase average of the difference between the inferred and input mean hemispheric fields ($\langle\langle B \rangle - \langle B \rangle_{\mathrm{input}} \rangle_p$), with the uncertainty corresponding to standard deviation of this difference. This metric serves as an indicator of systematic biases and discrepancies between the estimated and actual inference errors. The last three columns present the value of the maximum field strength ($B_\mathrm{max}$) corresponding to the number of magnetic components that minimised BIC, AIC, and WAIC respectively for the highest number of rotation phases. For cases in which different criteria selected different values of $B_\mathrm{max}$, the corresponding inferred \bs$_p$ and $\langle\langle B \rangle - \langle B \rangle_{\mathrm{input}} \rangle_p$ are listed in separate rows. 

\begin{table*}[t!]
\caption{Input parameters for field retrieval simulations, the derived phase-averaged mean surface field strength, its difference with respect to the input field, and the maximum field strengths according to different statistical criteria.}   
\label{tab:input_par}    
\centering                    
\begin{tabular}{c c c c c c c c c c }      
\hline\hline            
$s$ & \bs$_s$ & \vsini\ & $i$ & \bs$_{p,\mathrm{input}}$ & \bs$_p$ & $ \langle\langle B \rangle - \langle B \rangle_{\mathrm{input}} \rangle_p$ & $B_\mathrm{max}$(BIC) & $B_\mathrm{max}$(AIC) & $B_\mathrm{max}$(WAIC)\\
 & (kG) & (\kms) & (\degr) & (kG) & (kG) & (kG) & (kG) & (kG) & (kG) \\
\hline                      
   0.5 & 0.792 & 1 & 45 & 0.81 & 0.613$^{+0.034}_{-0.034}$ & $-0.198 \pm 0.038$ & 2 & 2 & 2\\    
       & & 3 & 45 & 0.81 & 0.705$^{+0.041}_{-0.041}$ & $-0.106 \pm 0.043$ & 2 & 2 & 2 \\ 
       & & 5 & 45 & 0.81 & 0.779$^{+0.050}_{-0.050}$ & $-0.032 \pm 0.036$ & 2 & 2 & 2\\ 
       \hline 
  1    & 1.584 & 1 & 45 & 1.62 & 1.30$^{+0.054}_{-0.053}$ & $-0.321 \pm 0.039$ & 6 & 6 & 6 \\   
       & & 3 & 45 & 1.62 & 1.40$^{+0.070}_{-0.070}$ & $-0.217 \pm 0.093$ & 6 & 6 & 6 \\ 
       & & 5 & 45 & 1.62 & 1.55$^{+0.088}_{-0.088}$ & $-0.074 \pm 0.071$ & 6 & 6 & 6 \\  
       & & 10 & 45 & 1.62 & 1.533$^{+0.095}_{-0.095}$  & $-0.09 \pm 0.10$ & 4 & 4 & --\\
       & &    &    &      & 1.61$^{+0.10}_{-0.11}$  & $-0.01 \pm 0.12$ & -- & -- & 6\tablefootmark{a} \\
       
       \hline
  2    & 3.168 & 3 & 45 & 3.24 & 2.97$^{+0.11}_{-0.10}$ & $-0.27 \pm 0.10$ & 12 & 12 & --\\
     & & & & & 3.08$^{+0.12}_{-0.12}$ & $-0.16 \pm 0.10$ & -- & -- & 14 \\
       & & 5 & 15 & 3.75 & 3.68$^{+0.13}_{-0.13}$ & $-0.07 \pm 0.11$ & 12 & 12 & 12 \\
       & &   & 45 & 3.24 & 3.15$^{+0.13}_{-0.13}$ & $-0.09 \pm 0.13$ & 12 & 12 & 12 \\ 
       & &   & 75 & 2.79 & 2.68$^{+0.13}_{-0.13}$ & $-0.11 \pm 0.19$ & 12 & 12 & 12 \\
       & & 10 & 45 & 3.24 & 3.06$^{+0.16}_{-0.16}$ & $-0.18 \pm 0.24$ & 10 & -- & -- \\
       & &  & &  & 3.24$^{+0.18}_{-0.18}$ & $-0.01 \pm 0.22$ & -- & 12 & 12 \\
       & & 20 & 45 & 3.24 & 2.72$^{+0.17}_{-0.17}$ & $-0.52 \pm 0.15$ & 6 & -- & --\\
       & &  &  &  & 2.97$^{+0.19}_{-0.19}$ & $-0.27 \pm 0.19$ & -- & 8 & --\\ 
       & &  &  &  & 3.26$^{+0.22}_{-0.21}$ & $0.02 \pm 0.17$ & -- & -- & 12\\ 
\hline                                 
\end{tabular}
\tablefoot{Uncertainties in column six correspond to phase-average of formal errors obtained from posterior distributions. Those in column seven represent the standard deviation of the difference between the inferred and input field strength. Results are reported separately in all cases when different statistical criteria preferred different $B_\mathrm{max}$.}\\
\tablefoottext{a}{For this case, WAIC was minimised by $B_\mathrm{max} = 4$ and 6~kG for the same number of rotation phases. 
}
\end{table*}

\subsection{Reference active dwarf}
\label{sub_sec:results_reference}

This section presents the results obtained from the analysis of simulated observations for the case $s=2$, \vsini\,=\,5~\kms\ and $i=45$\degr, corresponding to a relatively active M dwarf. We chose this case as reference for an in-depth discussion of the results.

For the determination of all parameters we adopted non-informative priors to guide the MCMC sampling without setting any specific input values, simulating typical inference from observations. The Ti abundance was given a uniform prior between $-6.95$ and $-7.15$ on the $\log N_{\rm Ti}/N_{\rm tot}$ scale. All magnetic filling factors had a uniform prior between 0 and 1, excluding the first factor, $f_0$, deriving from the normalisation constraint (Eq.~\ref{eq:filling_fac_norm}). For the case of $s=2$, \vsini\,=\,5~\kms, and $i=45$\degr, we set a uniform prior on the projected rotational velocity of $6 \pm 2.5$~\kms. Here, as well as in all cases described below, MCMC was initialised with equal values of magnetic filling factors and both $\epsilon(Ti)$ and \vsini\ offset from their respective true values.

The magnetic field was inferred using parametrisations with a different number of magnetic filling factors. The model chosen was the one with the lowest WAIC value, computed according to Eq.~(\ref{eq:WAIC}). For the example case, the chosen number of filling factors was 7, spanning values of magnetic fields from 0~kG to 12~kG. For each rotational phase, \vsini\, and $\epsilon(Ti)$ were inferred simultaneously with the magnetic field.  

\begin{figure*}[!t]
    \centering
    \includegraphics[width=\hsize]{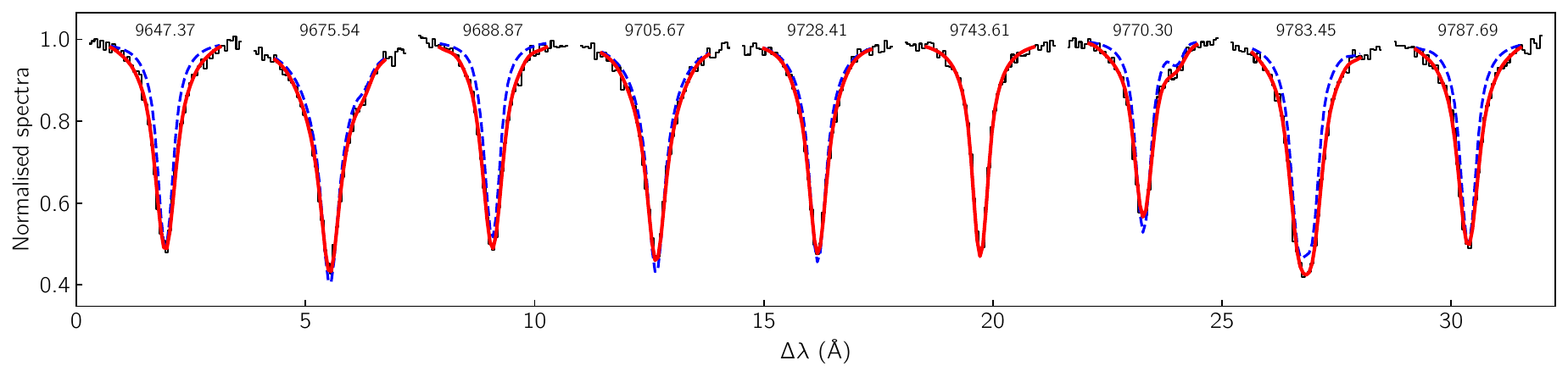}
    \caption{Comparison between simulated input spectra (black histograms) and model spectra corresponding to median parameters (solid red lines, corresponding to the central orange points in Fig.~\ref{fig:post_dist_s2_v5_i45_p0.0}) for the case of $s=2$, \vsini\,=\,5~\kms, $i=45$\degr, and rotational phase $p=0.0$. The dashed blue lines correspond to synthetic spectra without the magnetic field.}
    \label{fig:lines_s2_v5_i45_p0.0}
\end{figure*}

A detailed account of the results for phase $p=0.0$ is shown, as example, in Fig.~\ref{fig:post_dist_s2_v5_i45_p0.0} and Fig.~\ref{fig:lines_s2_v5_i45_p0.0}. Figure~\ref{fig:post_dist_s2_v5_i45_p0.0} illustrates, for this rotational phase, the posterior distribution of the inferred parameters: filling factors, hemispheric average magnetic field (\bs), Ti abundance, projected rotational velocity, and continuum scaling factors. The plots that are not on the diagonal show correlations between parameters. Filling factors of different magnetic components are correlated with each other. This is to be expected especially between neighbouring field components, as the spectral line profiles are primarily sensitive to the average field strength rather than to the detailed shape of the field strength distribution. Correlations between filling factors and \vsini\ are also present, due to the effect of the rotational broadening. 
On the other hand, the continuum scaling factors do not exhibit noticeable correlations between themselves or with other parameters and, for all lines, are recovered close to the input value of unity.
Figure~\ref{fig:lines_s2_v5_i45_p0.0} shows a comparison between input simulated spectrum and the model computed with the median parameters.

\begin{figure}
    \centering
    \includegraphics[width=1\linewidth]{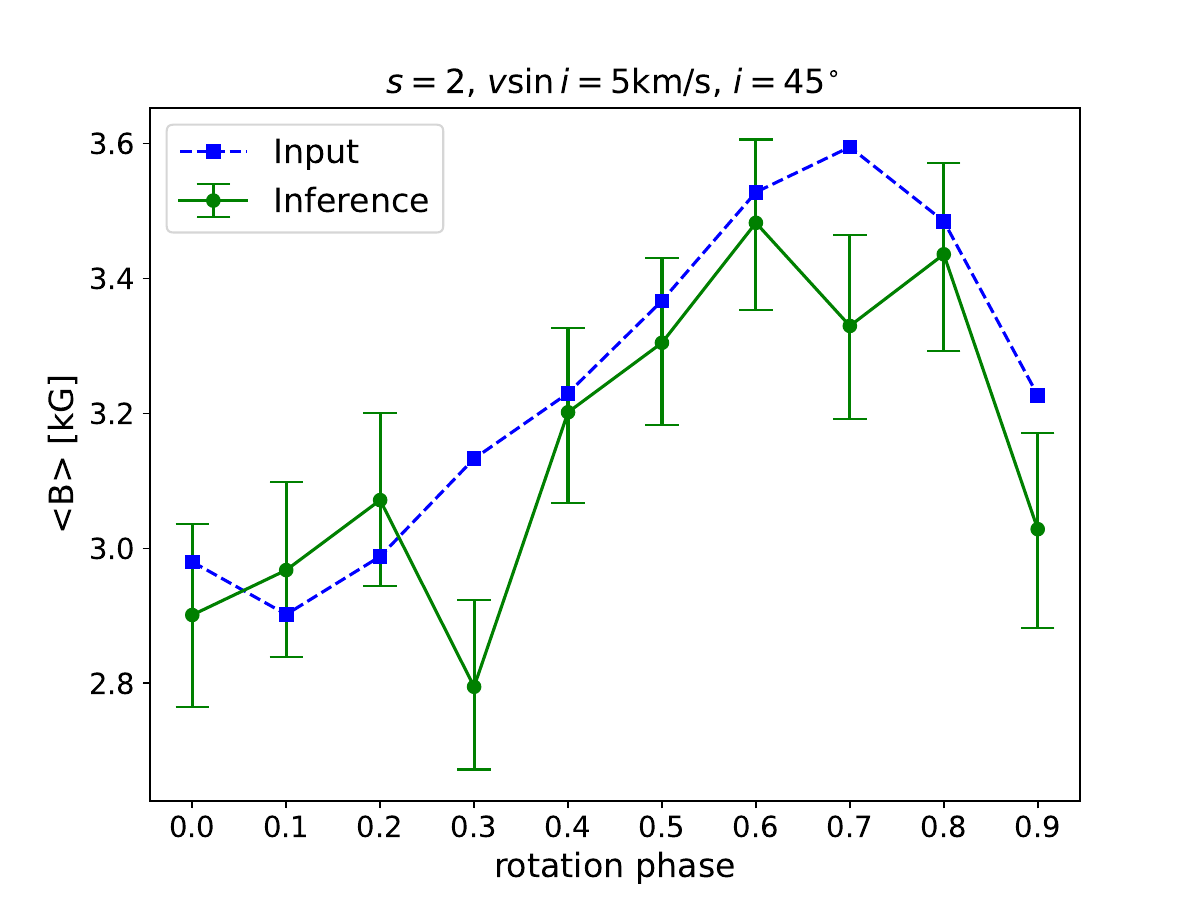}
    \caption{Inferred values of the hemispheric average magnetic field (solid green line) and \bs\ values calculated from the input magnetic map (dashed blue) for the case of $s=2$, \vsini\,=\,5~\kms, $i=45$\degr.}
    \label{fig:input_recB_s2_v5_i45}
\end{figure}

The inferred values of the hemispheric surface average magnetic field, \bs, are presented in Fig.~\ref{fig:input_recB_s2_v5_i45}\footnote{Equivalent figures for other test cases that are not discussed in Sect.~\ref{Sec:Results} are presented in Appendix~\ref{appen_2kG}.} with comparison to the values extracted from the input magnetic map. For the majority of rotational phases, the input and inferred values agree within the uncertainty. The hemispheric magnetic field averaged over the rotation phases is \bs$_p =  3.15 \pm 0.22$~kG, which is compatible with the $s=2$ scaled up version of the input MHD field structure model. The inferred value of the projected rotational velocity averaged over the phases is $5.12\pm 0.19$~\kms. Lastly, the mean Ti abundance is $-7.050\pm0.015$. Both are compatible with input values.

Results of the inference of the hemispheric surface average field, projected rotational velocity and Ti abundance for each rotational phase are summarised in Table~\ref{tab:results_s2_v5_i45}. For each parameter, the median values are reported alongside the 68\% confidence intervals. 

\begin{table}[]
    \centering
    \caption{Inferred parameters as a function of rotational phase $p$ for the case of $s=2$, \vsini\,= \,5~\kms, $i=45$\degr. }
    \begin{tabular}{c c c c}
    \hline \hline
       $p$ & \bs (kG) &  \vsini\ (\kms) & $\epsilon(Ti)$\\
       \hline
       0.0 & $2.90^{+0.14}_{-0.14}$  & $5.11^{+0.17}_{-0.16}$ & $-7.020^{+0.015}_{-0.015}$ \\
       0.1 & $2.97^{+0.13}_{-0.13}$ & $4.91^{+0.17}_{-0.16}$ & $-7.067^{+0.014}_{-0.014}$ \\
       0.2 & $3.07^{+0.13}_{-0.13}$ & $5.04^{+0.18}_{-0.18}$ & $-7.038^{+0.013}_{-0.014}$ \\
       0.3 & $2.79^{+0.13}_{-0.12}$ & $5.29^{+0.17}_{-0.16}$ & $-7.057^{+0.015}_{-0.014}$ \\
       0.4 & $3.20^{+0.12}_{-0.13}$ & $5.29^{+0.16}_{-0.16}$ & $-7.053^{+0.014}_{-0.014}$ \\
       0.5 & $3.30^{+0.13}_{-0.12}$ & $4.89^{+0.18}_{-0.18}$ & $-7.070^{+0.014}_{-0.015}$ \\
       0.6 & $3.48^{+0.12}_{-0.13}$ & $4.96^{+0.18}_{-0.18}$ & $-7.051^{+0.014}_{-0.013}$ \\
       0.7 & $3.33^{+0.13}_{-0.14}$ & $5.18^{+0.18}_{-0.18}$ & $-7.062^{+0.014}_{-0.014}$ \\
       0.8 & $3.44^{+0.14}_{-0.14}$ & $5.20^{+0.18}_{-0.19}$ & $-7.046^{+0.015}_{-0.014}$ \\
       0.9 & $3.03^{+0.14}_{-0.15}$ & $5.52^{+0.18}_{-0.18}$ & $-7.031^{+0.015}_{-0.015}$ \\
       \hline
    \end{tabular}
    
    \label{tab:results_s2_v5_i45}
\end{table}

\subsection{Choice of maximum field strength}
\label{sub_sec:results_choice_max_field}

An essential step in the inference of the magnetic field distribution is the choice of the number of magnetic filling factors. Various approaches have been proposed in the literature, as outlined in Sect.~\ref{sub_sec:Inference}. To evaluate this choice in our framework, we applied models with different numbers of magnetic components across a set of representative input parameters and computed the BIC, AIC, and WAIC for each case. The tested cases and the corresponding maximum field strength favoured by each criterion are summarised in Table~\ref{tab:input_par} and are discussed in this section. We found that the three criteria were generally minimised by the same number of components for simulated stars with lower magnetic activity and slow rotation. In contrast, criteria diverged for simulated stars with stronger magnetic activity, and, as expected, they generally favoured models with more magnetic components, and therefore a higher maximum magnetic field value.

For $s=0.5$, which corresponds to a surface average input magnetic field of \bs$_s$\,$ \approx$\,0.8~kG, all criteria agreed on two magnetic components, with field strengths of 0 and 2~kG, for the three considered projected rotational velocities (1, 3, and 5~\kms) and $i=45$\degr. 

In the case of $s=1$, with an average surface field of \bs$_s$\,$\approx$\,1.6~kG, for velocities between 1 and 5~\kms, a 4-components model, corresponding to a maximum value of magnetic field of $B_\mathrm{max}=6$~kG, minimised all criteria. 

For the less active and slower rotating cases, namely $s=0.5$, 1 and \vsini\,=\,1 and 3~\kms, we observed an underestimation of \bs\ by 13--24\%, as documented in Table~\ref{tab:input_par} and illustrated in Figs.~\ref{fig:input_recB_s05_i45}--\ref{fig:input_recB_s1_i45}. This bias gradually disappeared as \vsini\ increased beyond 3~\kms.

\begin{figure*}
    \centering
    \includegraphics[scale=0.45]{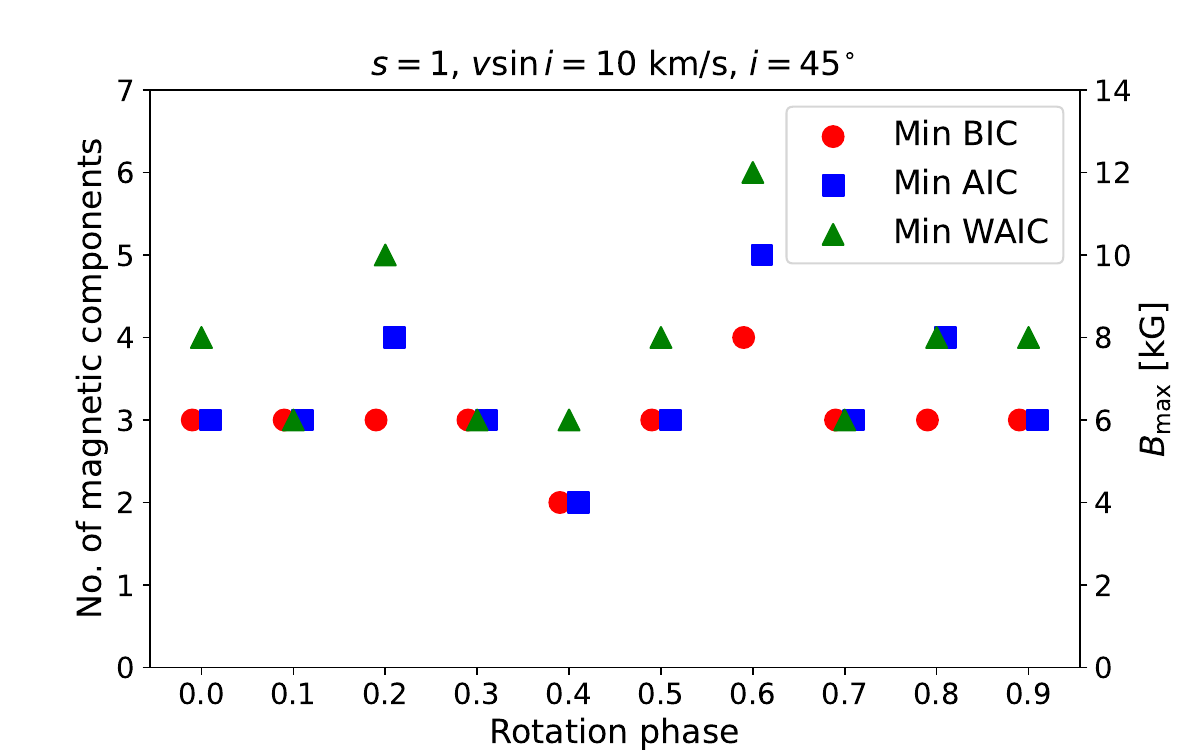}
    \includegraphics[scale=0.45]{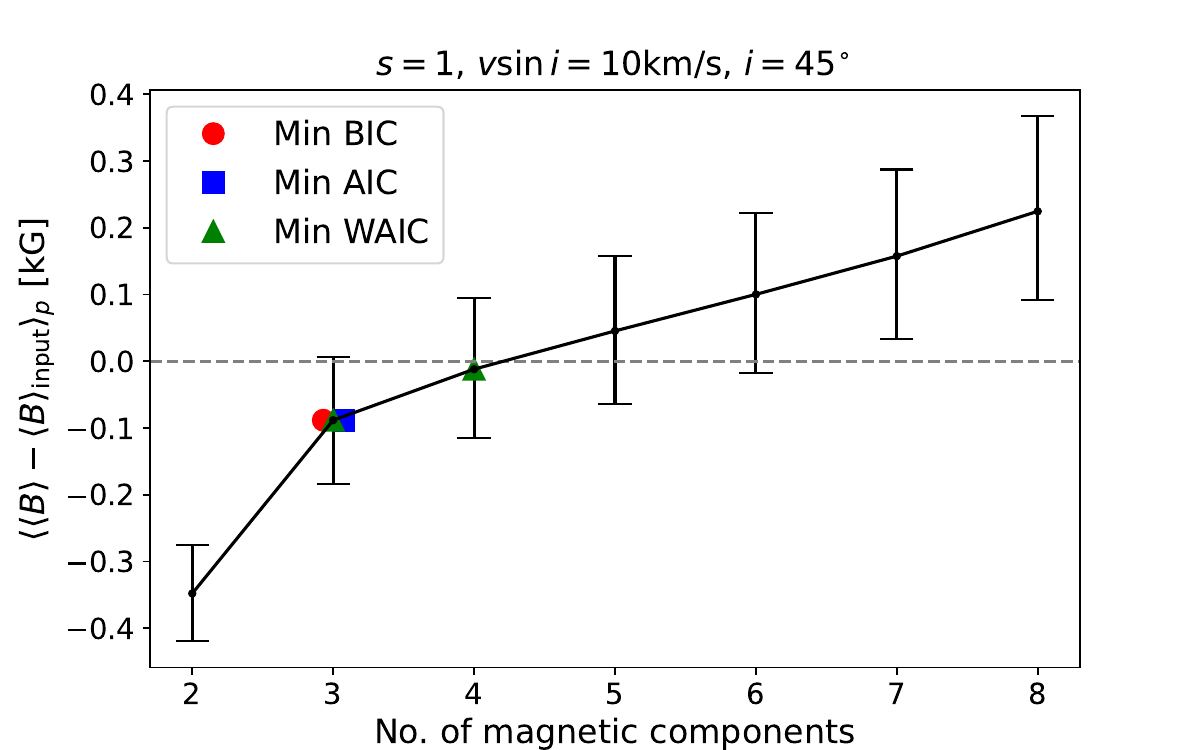}
    \caption{Results of the inference with the number of magnetic field components minimised by BIC (red circles), AIC (blue squares), and WAIC (green triangles) for the case of $s=1$, \vsini\,=\,10 \kms, $i=45$\degr. \textit{Left}: Number of components that minimised BIC, AIC, and WAIC for each rotation phase. \textit{Right}: Difference, averaged over rotation phases, between the hemispheric average magnetic field, inferred from the number of magnetic components favouring each criterion, and the input field. For WAIC, two markers (3 and 4 components) are shown, as these models were favoured by the same number of rotation phases.}
    \label{fig:criteria_diff_s1_v10_i45}
\end{figure*}

For a higher velocity of \vsini\,=\,10~\kms,  the criteria governing $B_{\rm max}$ selection diverged somewhat. BIC and AIC were minimised by a 3-component model, while WAIC did not identify a single preferred number of components, with 3- and 4-component models being both selected by four rotation phases, as shown in the left panel of Fig.~\ref{fig:criteria_diff_s1_v10_i45}. The right panel of this figure shows the difference, averaged over rotation phases, between the hemispheric surface average input magnetic field and the inferred field computed using models with different numbers of magnetic components. We concluded that the 4-component model, which minimised WAIC but not BIC or AIC, is the configuration that also yielded the smallest averaged difference between the input magnetic field and the inferred field, suggesting that WAIC may better capture subtle improvements in model performance in this regime. 

To assess how this result is affected by random noise added to simulated observations, we generated a new set of simulated line profiles with the same setup ($s=1$, \vsini\,=\,10~\kms, $i=45\degr$) but with a different noise realisation for each phase. We repeated the full inference and model selection analysis. These calculations\footnote{A detailed presentation of this analysis is included Appendix~\ref{appen_different-noise}.} showed that a phase-by-phase pattern of the preferred number of components differed by $\pm1$ from that shown in Fig.~\ref{fig:criteria_diff_s1_v10_i45} left panel. However, BIC and AIC continued to favour a 3-component model, while WAIC favoured a 4-component model, with only a small number of phases selecting 3 or 5 components. Importantly, the model favoured by WAIC again provided a better retrieval of $\langle B \rangle_p$, whereas the BIC and AIC-selected models underestimated it. This test demonstrates that, under our setup, random noise introduces some variability in the model selection at individual rotation phases. Nevertheless, the overall conclusions remain robust when based on phase-averaged results.

\begin{figure*}
    \centering
    \includegraphics[scale=0.45]{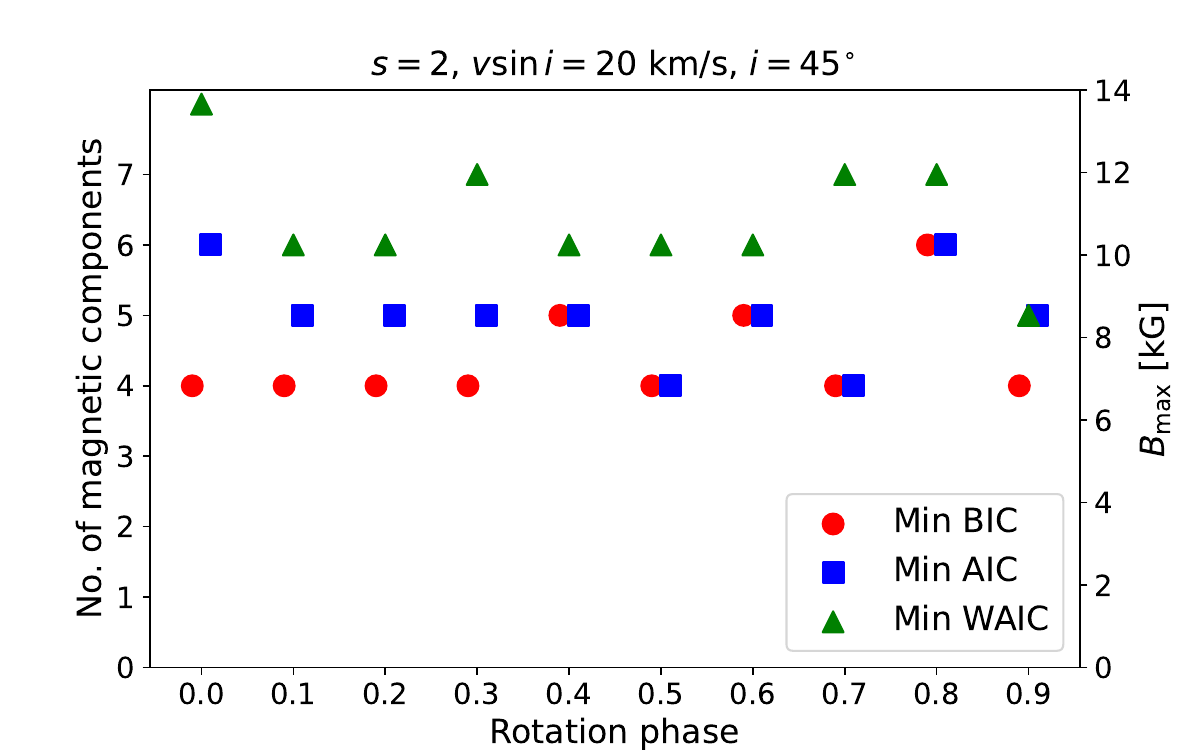}
    \includegraphics[scale=0.45]{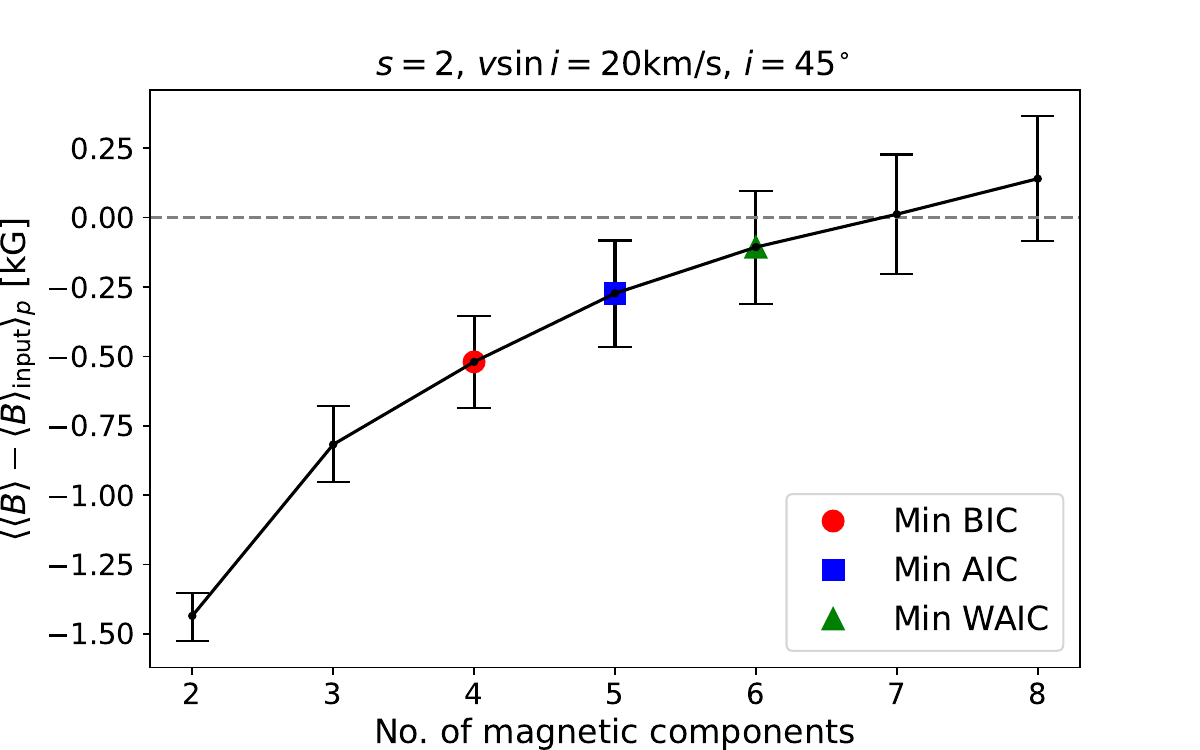}
    \caption{Same as Fig.~\ref{fig:criteria_diff_s1_v10_i45} but for the case of $s=2$, \vsini\,=\,20~\kms, $i=45$\degr. }
    \label{fig:criteria_diff_s2_v20_i45}    
\end{figure*}

Different criteria favoured different numbers of magnetic components also when analysing simulated observations with $s=2$, corresponding to a full-surface average input field of $\approx$\,3.2~kG. For \vsini\,=\,3~\kms\ and $i=45$\degr, BIC and AIC agreed on a 7-component model (corresponding to a maximum value of magnetic field of 12~kG), while WAIC was mostly minimised by a 8-components model (maximum magnetic field of 14~kG).
For the case of $s=2$, \vsini\,=\,5~\kms\ and all inclinations, all criteria were minimised by a 7-component model. 

At higher rotation speeds, discrepancies between criteria were observed.  For \vsini\,=\,10~\kms\ and $i=45$\degr, BIC favoured 6 magnetic components ($B_\mathrm{max}=10$~kG), while AIC and WAIC preferred 7 components ($B_\mathrm{max}=12$~kG). Lastly, for \vsini\,=\,20~\kms\ and $i=45$\degr, presented in Fig.~\ref{fig:criteria_diff_s2_v20_i45}, BIC favoured 4 magnetic components ($B_\mathrm{max}=6$~kG), AIC 5 components ($B_\mathrm{max}=8$~kG), and WAIC 6 components ($B_\mathrm{max}=10$~kG). However, it should be noted that for four rotation phases, WAIC was minimised by models with a higher number of components, yielding $B_\mathrm{max}$ values up to 14~kG. The right panel of Fig.~\ref{fig:criteria_diff_s2_v20_i45} shows the difference between the inferred field and input field for the latter experiment ($s=2$, \vsini\,=\,20~\kms), averaged over rotation phases, as a function of the number of magnetic components. It is evident that the configurations favoured by BIC and AIC tended to underestimate the magnetic field. The 6-component model selected by WAIC substantially reduces the discrepancy. Moreover, using seven components, which minimised WAIC in three out of ten rotation phases, resulted in an inferred field that was significantly closer to the true value obtained from the input map.

From this analysis of fast rotators, we concluded that selecting models with a higher number of magnetic components can be justified even if they are not consistently preferred across all rotation phases. Preference by WAIC in only a subset of phases may still indicate that these models provide a more accurate recovery of the hemispheric surface average magnetic field.

\subsection{Dependence on inclination}
\label{sub_sec:results_inclination}

In this section, we present the investigation of how different inclination angles influence the reconstruction of the magnetic field. As shown in Fig.~\ref{fig:MHD}, the input MHD model exhibits a systematic latitudinal variation, with stronger fields concentrated near the rotational poles. Consequently, observations of the star at different inclination angles sample distinct field-strength distributions and correspond to different average magnetic field values. It is not immediately clear, however, whether certain inclination angles introduce additional challenges for magnetic inference.

\begin{figure}
    \centering
    \includegraphics[width=1\linewidth]{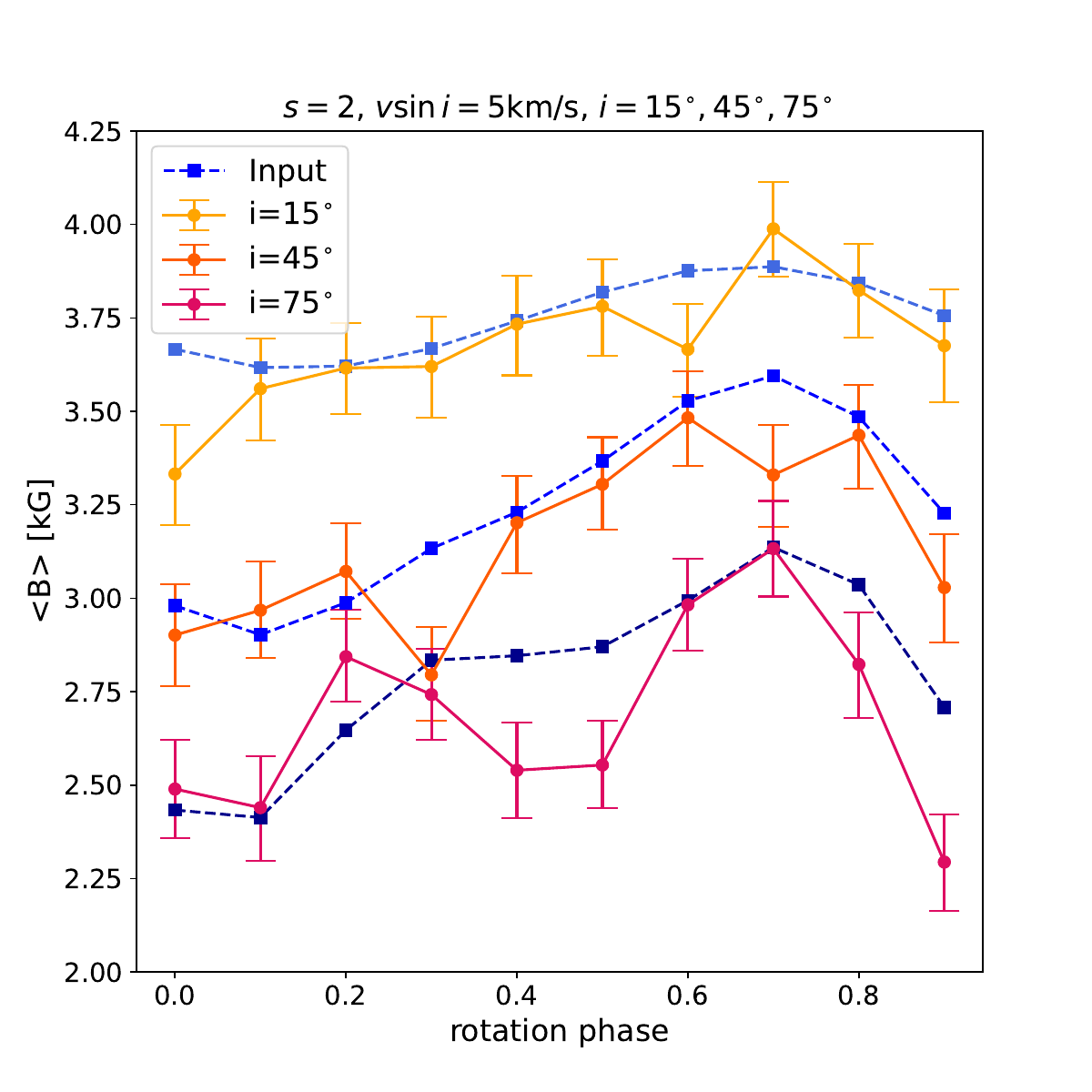}
    \caption{Inferred values of the hemispheric average magnetic field (solid lines) and values corresponding to the input map (dashed lines) for the case of $s=2$, \vsini\,=\,5~\kms, $i$\,=\,15\degr\ (top lines), 45\degr\ (middle lines), and 75\degr\ (bottom lines).}
    \label{fig:diff_theo_inf_s2_v5_all_i}
\end{figure}

For this investigation, we focused on the reference case of $s=2$ and \vsini\,=\,5~\kms, and analysed three inclinations: $i=15^{\circ}$, $45^{\circ}$, and $75^{\circ}$. The reconstruction for $i=45^{\circ}$ corresponds to the reference case presented in Sect.~\ref{sub_sec:results_reference}, while the cases of $i=15^{\circ}$ and $75^{\circ}$ were treated in the same way. Based on the minimum WAIC, we adopted a model with seven magnetic filling factors for all inclinations, corresponding to a maximum surface field strength of $B_{\rm max}=12$~kG. Figure~\ref{fig:diff_theo_inf_s2_v5_all_i} compares the hemispheric averaged input magnetic field with the reconstructed field for the three inclinations. The largest divergence from the input field is seen for $i=75^{\circ}$, while the agreement is noticeably better at $i=45^{\circ}$ and especially at $i=15^{\circ}$.

This inclination dependence of the field strength inference results is consistent with the underlying distribution of the input field in the MHD model (Fig.~\ref{fig:MHD}), where magnetic flux is concentrated towards the poles. This is in contrast with the assumption of a uniform distribution assumed when constructing the synthetic spectrum. Due to limb darkening, areas on the edge of the stellar disk contribute less to the total flux. Thus, if most magnetic flux is produced in the polar regions, the corresponding signal is weaker and may be interpreted as a reduced magnetic filling factor. When the star is seen nearly pole-on ($i=15^{\circ}$), these polar concentrations dominate the visible hemisphere, and the reconstruction is more faithful to the input field. At higher inclinations, the polar regions contribute less to the disk-integrated spectrum, while equatorial regions with weaker fields contribute more, which may reduce the accuracy of the reconstruction, leading to underestimations. Therefore, the inclination angle can play an important role in determining how well small-scale spatially inhomogeneous fields are recovered.

\begin{figure*}[!th]
    \centering
    \includegraphics[width=1\linewidth]{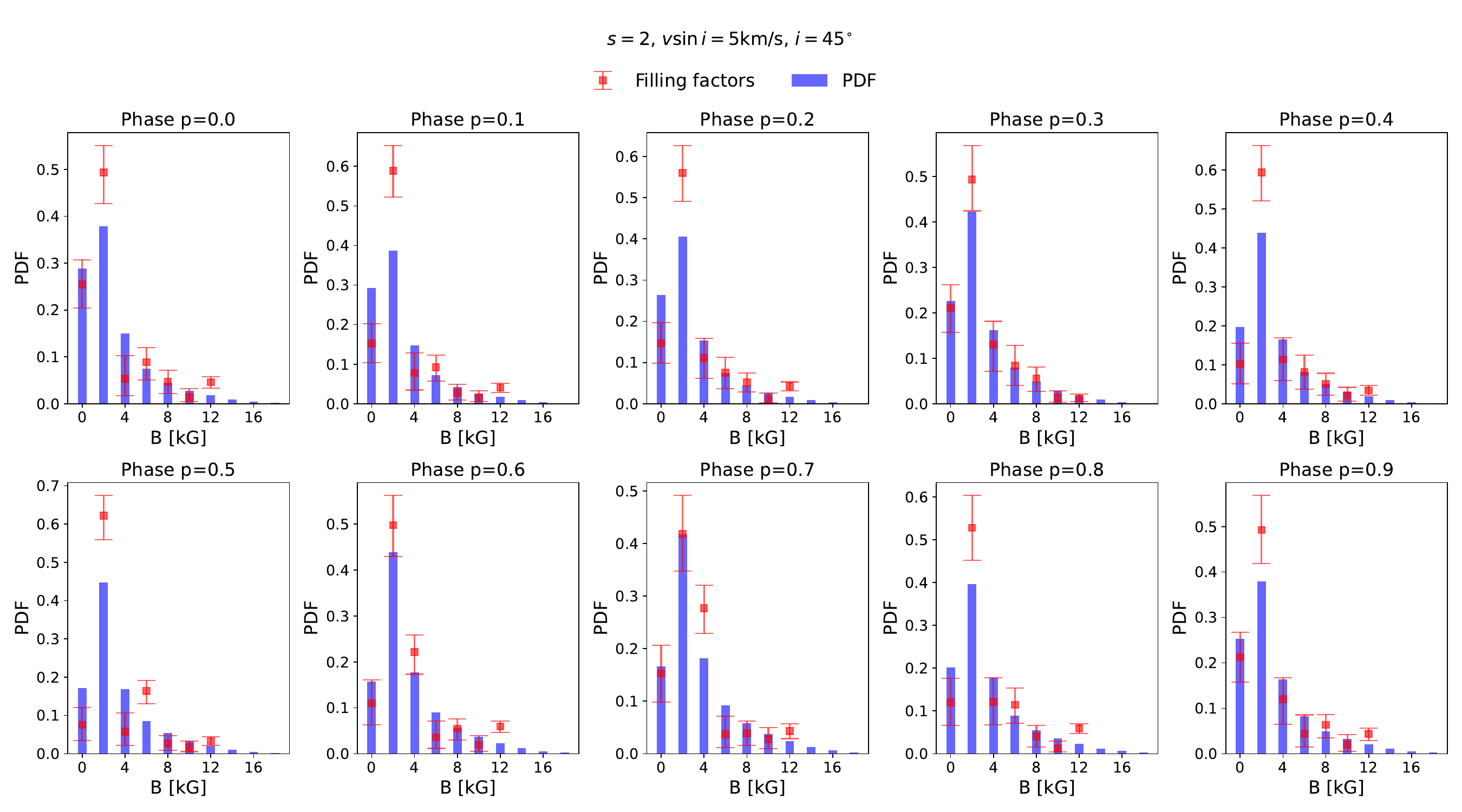}
    \caption{Inferred distributions of magnetic filling factors (red symbols with error bars) and corresponding input probability distribution function (PDF, blue histogram) for the case $s=2$, \vsini\,=\,5~\kms, $i=45$\degr, and ten rotation phases. The number of magnetic filling factors is chosen in accordance with WAIC. The PDF is truncated at 18~kG. }
    \label{fig:filling_fact_s2_v5_i45}
\end{figure*}

\subsection{Recovery of field strength distributions}

In Sect.~\ref{sub_sec:results_choice_max_field} we discussed the choice of the number of magnetic components. Each component is associated with a filling factor, which is a free parameter describing the fractional surface coverage of that field strength. Together, the inferred filling factors provide an approximation of the distribution of magnetic field strength over the visible part of the stellar surface. It is of interest to examine to what extent the set of filling factors obtained in our inferences reproduces the actual field strength distribution in the relevant part of the input MHD map.

An example of this assessment is shown in Fig.~\ref{fig:filling_fact_s2_v5_i45} for the case of $s=2$, \vsini\,=\,5~\kms, $i=45$\degr. The plot compares the inferred magnetic filling factors (red squares) with the input probability distribution function (PDF, blue histogram), which was computed by weighting the local field strengths with the surface area of each grid element, its visibility to the observer, and a limb-darkening factor. The PDF was then normalised and arbitrarily truncated at 18\,kG, as only a negligible fraction of the surface exhibits magnetic fields of such high strength. The inference model included seven magnetic components, as favoured by WAIC.

Overall, the reconstruction reproduces the general shape of the input field strength distribution, correctly capturing both the location of the peak and the steep decline towards higher field strengths with filling factors deviating from the input field PDF by 1.3$\sigma$ on average. This metric corresponds to the average, over magnetic components and rotation phases, of the absolute differences between inferred filling factors and the input binned PDF, normalised by the corresponding uncertainties.

However, some systematic discrepancies are visible. In particular, the contribution of the second magnetic component is overestimated, while the contribution of the first is slightly underestimated. Lastly, the contribution of the 12~kG component is slightly overestimated, likely because the remaining high-field tail of the distribution is absorbed into this final bin. This trend was observed for most analysed cases. These results suggest that, although the multicomponent model reproduces the overall field distribution well, the discretisation into a finite number of components inevitably redistributes part of the signal.

\subsection{Alternative fitting strategy}
\label{sub_sec:results_comparisonR22}

As discussed in Sect.~\ref{sub_sec:Inference}, in our inference framework, the continuum around each spectral line is treated as a free multiplicative scaling factor, the relative line strengths are fixed according to adopted line list parameters, and the Ti abundance is allowed to vary. This standard approach, adopted by many observational studies, requires accurate input atomic data but is applicable to both slow and fast rotators since it utilises information from both line shapes (differential Zeeman broadening) and line intensities (differential magnetic intensification).
 
A different strategy was adopted by \citet{Reiners2022} (hereafter R22). In their approach, the continuum level and elemental abundances were fixed, while line-strength scaling factors were fitted as part of the inference. This procedure allowed the relative depths of individual lines to be adjusted according to an optical-depth scaling law, thereby compensating for discrepancies caused by uncertainties in oscillator strengths, stellar parameters, and abundances. Although this method offers certain computational advantages over the standard approach, it entirely forgoes the use of magnetic intensification to constrain magnetic field properties and is therefore susceptible to additional biases. 

In this section, we present a comparison between the standard inference approach and the R22 method. To make the comparison between the two methods more realistic, we generated a new set of simulated observations to match the setup of R22. We set a resolution of $R=80\,400$, equal to the resolution of the NIR channel of the CARMENES spectrometer used by R22, adopt 1.5~\kms\, step between pixels, equal to the step in CARMENES spectra, and use a signal-to-noise ratio S/N=300, typical of the observational data analysed by R22. We set input parameters to test the cases of $s=0.5$, \vsini\,=\,1, 3, 5~\kms, and $i=45\degr$.

To match the analysis presented in R22, the magnetic field was parametrised in steps of 1~kG, with a maximum magnetic field value of 4~kG. This arbitrary choice replaces our quantitative statistical approach used elsewhere in this paper to determine $B_{\rm max}$. The simulated spectra for the mentioned test cases were then processed with the two inference methods that differ in the treatment of the spectral line strengths: continuum scaling around spectral lines and Ti abundance treated as free parameters (standard method) and line strength scaling approach (R22 method).

\begin{figure} 
    \centering
    \includegraphics[width=1\linewidth]{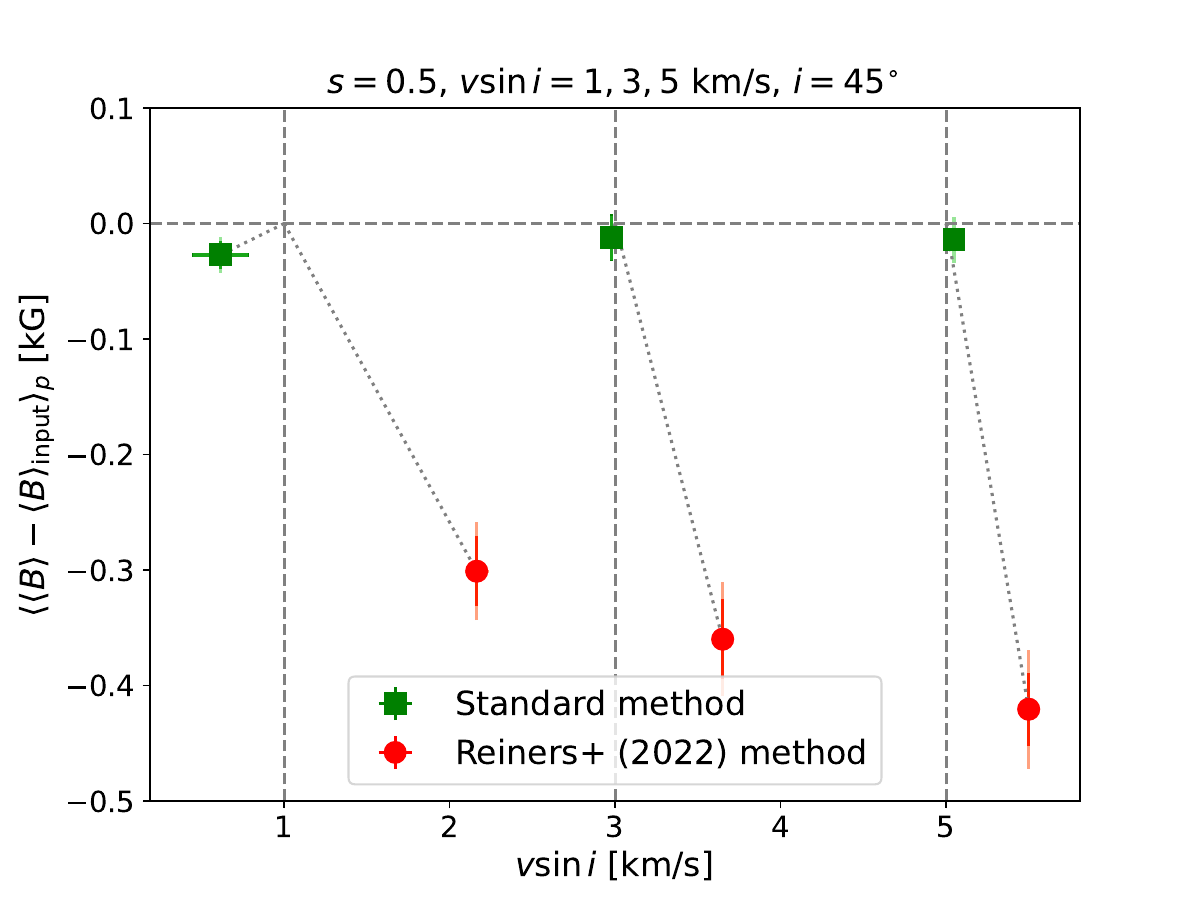}
    \caption{Difference, averaged over the rotation phases, between the hemispheric-surface-average inferred magnetic field and the input field as a function of the inferred projected rotational velocity for the input cases $s=0.5$, \vsini\,=\,1, 3, 5~\kms, $i=45\degr$. The green squares are the result of the inference with continuum scaling factors as free parameters. The red circles are the results of the inference using \cite{Reiners2022} method with scaling of individual line strengths. Each point presents two types of error: standard deviation of the values across the rotation phases being shown by darker bars, and average of inferred errors in lighter bars.}
    \label{fig:comparisonR22_s0.5}
\end{figure}

Results of the comparison are summarised in Table~\ref{tab:results-R22}, presenting the input and inferred magnetic field strength parameters from the standard inference method and the R22 method, analogously to Table~\ref{tab:input_par}. Additionally, Fig.~\ref{fig:comparisonR22_s0.5} shows the difference between the inferred magnetic field and the input field, averaged over the rotation phases, ($ \langle\langle B \rangle - \langle B \rangle_{\mathrm{input}} \rangle_p$) as a function of \vsini. Appendix~\ref{appen_1kG} presents additional figures showing the inferred phase-resolved field strengths obtained with the standard inference method and with the R22 inference method, together with the corresponding values extracted from the input magnetic map.

We observed that the R22 method systematically underestimates the average magnetic field strength by $\sim$30--50\% relative to the true values, and at the same time tends to overestimate \vsini. This bias significantly exceeds the formal uncertainties. We therefore conclude that fitting the line-strength scaling factors, while helpful for compensating systematic errors in spectrum synthesis, can introduce a systematic underestimation of the hemispheric average magnetic field strengths. This bias increases with \vsini, reflecting gradual loss of magnetic line profile information due to Doppler broadening.

Lastly, results derived here using the standard inference method with $B_{\rm max}=4$~kG and 1~kG step demonstrate a better retrieval of the input field when compared (see Figs.~\ref{fig:input_recB_s05_i45} and \ref{fig:input_recB_s05_i45_R22_ourMethod}) with the earlier calculations in Sect.~\ref{sub_sec:results_choice_max_field}, which adopted $B_{\rm max}=2$~kG and 2~kG step. Most notably for $s=0.5$, \vsini\,=\,1, 3~\kms, the phase-averaged field strength underestimation reduced from 13--24\% to below 5\%. To assess in detail the effects of step size and choice of number of field components, for the \vsini\,=\,1~\kms\ case, we inspected the inference restricting $B_\mathrm{max}$ to 2~kG and 3~kG. Adopting a $B_\mathrm{max}=2$~kG model led again to an underestimation of the average magnetic field, similarly to the 2~kG grid inference. Conversely, setting $B_\mathrm{max}=3$~kG, which in this case was favoured by an analysis on BIC, AIC, and WAIC, led to a significantly improved reconstruction, while extending the grid further to 4~kG did not lead to additional improvement. Additionally, for the 2~kG binning, the inclusion of a higher field strength component ($B_\mathrm{max}=4$~kG) -- in this case not supported by any statistical criteria -- also did not improve the inference, with the highest-field component contributing negligibly to the derived filling factors. The results of these tests are summarised in Table~\ref{tab:test-step-size-R22}. This analysis highlights the role of the magnetic field binning in achieving an accurate reconstruction, with WAIC-informed $B_{\rm max}$ still providing an adequate inference.

\begin{table*}[t!]
\caption{Input parameters for field retrieval simulations, derived phase-averaged mean surface field strength, and its difference with respect to the input field for the weak-activity case discussed in Sect.~\ref{sub_sec:results_comparisonR22}.}  
\label{tab:results-R22}    
\centering           
\begin{tabular}{c c | c c | c c}      
\hline\hline            
& & \multicolumn{2}{c|}{Standard method} & \multicolumn{2}{c}{\citet{Reiners2022} method} \\
 \vsini\ & \bs$_{p,\mathrm{input}}$ & \bs$_p$ & $ \langle\langle B \rangle - \langle B \rangle_{\mathrm{input}} \rangle_p$ & \bs$_p$ & $ \langle\langle B \rangle - \langle B \rangle_{\mathrm{input}}  \rangle_p$ \\
 (\kms) & (kG) & (kG) & (kG) & (kG) & (kG) \\
\hline                      
   1 & 0.81 & 0.790$^{+0.016}_{-0.016}$ & $-0.021\pm 0.012$ & 0.510$^{+0.042}_{-0.043}$ & $-0.301\pm0.030$ \\
   3 & 0.81 & 0.799$^{+0.018}_{-0.020}$ & $-0.012\pm 0.020$ & 0.451$^{+0.051}_{-0.047}$ & $-0.360\pm0.034$ \\
   5 & 0.81 & 0.797$^{+0.020}_{-0.019}$ & $-0.0141\pm 0.0070$ & 0.390$^{+0.054}_{-0.049}$ & $-0.421\pm0.032$ \\
\hline                                 
\end{tabular}

\end{table*}

\section{Summary and conclusions}
\label{sec:Conclusions}

This work has presented an analysis of commonly employed Zeeman broadening techniques for measuring magnetic fields in M dwarfs. Using the high-resolution 3D MHD model by \citet{Yadav2015} combined with MARCS model atmospheres \citep{Gustafsson2008}, we generated synthetic Stokes~$I$ spectra of ten \ion{Ti}{i} lines in the $\lambda$ 964.74–978.77~nm range. The spectra were convolved with an instrumental profile ($R=65\,000$, velocity bin size of 2.0~\kms) and degraded to S/N = 100 by adding Gaussian noise. To simulate stars with different magnetic activity levels, the MHD model was scaled by factors of $s=0.5$, 1, and 2, corresponding to the surface average fields of $\sim$\,0.8, 1.6, and 3.2~kG. Additional sets of simulated observations were produced for a range of projected rotational velocities and inclination angles, as summarised in Table~\ref{tab:input_par}.

For the inference, we assumed a uniform radial magnetic field and parametrised the surface field using discrete filling factors in steps of 2~kG, normalised to unity. Each spectral line was assigned a multiplicative scaling factor to adjust its continuum, while Ti abundance was also treated as a free parameter. Stellar and magnetic parameters were estimated through MCMC sampling with the SoBAT library for IDL \citep{Anfinogentov2021}, as modified by \citet{Hahlin2022,Hahlin2023}.

A central aspect of the reconstruction was the choice of the number of magnetic filling factors. We evaluated this using three statistical criteria: the Bayesian Information Criterion (BIC), the Akaike Information Criterion (AIC), and the Widely Applicable Information Criterion (WAIC). We compared the resulting hemispheric averaged magnetic fields with the input values to assess accuracy. We found that the same number of magnetic components minimised all three criteria for the majority of the considered cases but diverged when considering more active and fast rotating stars. In those cases, BIC and AIC generally favoured fewer components and tended to produce lower inferred field strengths than WAIC. In contrast, WAIC typically selected models with more components, in some cases yielding reconstructions that more closely reproduced the input fields. Furthermore, when WAIC indicated additional components for only a subset of rotational phases, adopting the higher number of components generally reduced the discrepancy between the input and the inferred field strengths. 

However, even using a higher maximum field strength favoured by WAIC, we found situations when the application of the standard inference approach with the commonly adopted 2~kG field strength step yielded systematically biased inference results. In particular, the field strength retrieved for slowly rotating, weakly active stars was underestimated by up to 24\%. This bias was considerably less pronounced for both more rapidly rotating inactive stars and for all test cases corresponding to active M dwarfs. Additional numerical tests suggested that this field strength underestimation for inactive slow rotators disappears in the inference based on data with higher resolving power and S/N and finer filling factor sampling.

We then examined the influence of the inclination by comparing results for the same simulated star observed at different inclinations. We found that reconstructions for stars observed closer to the poles reproduced the input fields more reliably than for equator-on orientations, consistent with the polar concentration of magnetic flux in the underlying MHD model. Increasing the inclination generally reduced reconstruction accuracy and introduced larger phase-to-phase fluctuations.

We also investigated the recovery of the magnetic field strength distribution through the inferred filling factors. The overall shape of the input probability distribution function was reproduced, including the location of the peak and the decline towards higher field strengths. Nonetheless, systematic trends were present, with signal redistributed between neighbouring components due to the coarse discretisation. Additionally, a comparison with the results presented in Appendix~\ref{appen_1kG} shows that, for slowly rotating stars observed with higher resolution and higher S/N, adopting a finer magnetic field binning leads to a closer agreement between inferred and input hemispheric average field strengths.

Finally, we compared the standard line strength fitting strategy with the alternative method by \citet{Reiners2022}, in which individual line strengths are scaled freely rather than adjusted by modifying element abundance. After producing simulated observations with quality comparable to the observations analysed in their work, we concluded that the R22 method systematically underestimated the average surface magnetic field by $\sim$30–50\% and tended to overestimate \vsini.

Our results demonstrate that the Zeeman broadening diagnostic can provide a reliable reconstruction of the average magnetic fields in M dwarfs, but its performance is strongly influenced by methodological choices. In particular, the selection of the number of magnetic components and the adopted fitting strategy have a clear impact on the accuracy of the inferred field. Our analysis highlights the importance of guiding the choice of the number of magnetic components with statistical criteria. Furthermore, fitting line strengths instead of the continuum results in a systematic underestimation of the field. 

This study was designed to evaluate the accuracy of typical Zeeman broadening analyses of M dwarfs reported in the literature. To mirror the approach adopted in most previous studies, we employed a fixed magnetic field binning of 2~kG. However, our tests on weakly magnetic M dwarfs using observations with higher resolving power and S/N suggest that the chosen binning can influence the inferred field strength. This implies that the selection of an appropriate grid step warrants careful consideration.

Overall, our findings indicate that while the Zeeman diagnostic itself is robust, its effectiveness depends on how it is implemented. In future work, we will apply the best-performing strategies identified here to real high-resolution observations of M dwarfs. This development will contribute to building a consistent framework for magnetic field measurements in M dwarfs, with implications for both stellar physics and the characterisation of their planetary environments.

\begin{acknowledgements}
IA and OK acknowledge support by the Swedish Research Council (grant agreement no. 2023-03667). Additionally, OK acknowledges support by the Swedish National Space Agency. The authors wish to thank Dr. R. Yadav for making available his 3D MHD model of the fully convective M dwarf.
\end{acknowledgements}

\bibliographystyle{aa} 
\bibliography{references} 

\begin{appendix}

\onecolumn

\section{Example of posterior distributions}

This plot shows an example of posterior distributions derived with MCMC.

\begin{figure*}[!h]
    \centering
    \includegraphics[width=\hsize]{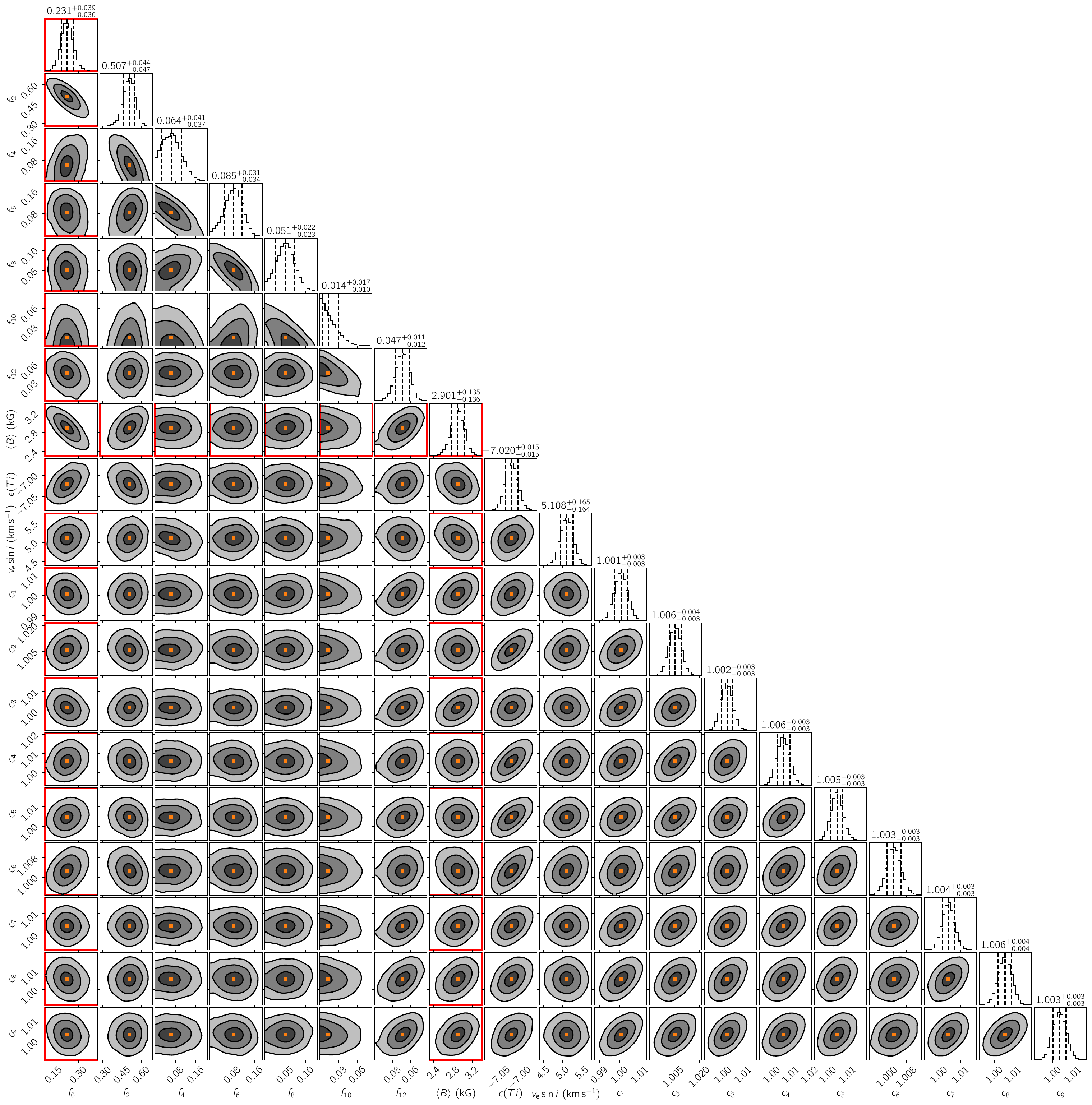}
    \caption{Posterior distributions for the case of $s=2$, \vsini\,=\,5~\kms, $i=45\degr$, and rotational phase $p=0.0$ showing the posterior distributions for the independent ($f_2$--$f_{12}$, $\epsilon (Ti)$, \vsini, $c_1$--$c_9$) and derived (\bs, $f_0$) parameters. The orange point in each panel represents the median value, while contours enclose regions of equal posterior probability density corresponding to 68, 95, and 99.7\% credible regions. The red borders highlight panels showing derived parameters.}
    \label{fig:post_dist_s2_v5_i45_p0.0}
\end{figure*}

\section{Inferred and input hemispheric average magnetic field}
\label{appen1}

\subsection{Standard inference}
\label{appen_2kG}

The following plots present the inferred values of the hemispheric surface average magnetic field, \bs, with comparison to the values extracted from the input magnetic map (dashed blue line). The inference employed magnetic filling factors defined in steps of 2~kG and different number of magnetic components minimising BIC (red line), AIC (blue), and WAIC (green). When only one inference line is shown, all criteria were minimised by the same number of magnetic components (see Sect.~\ref{sub_sec:results_choice_max_field}).

\begin{figure*}[!h]
    \centering
    \includegraphics[width=0.33\linewidth]{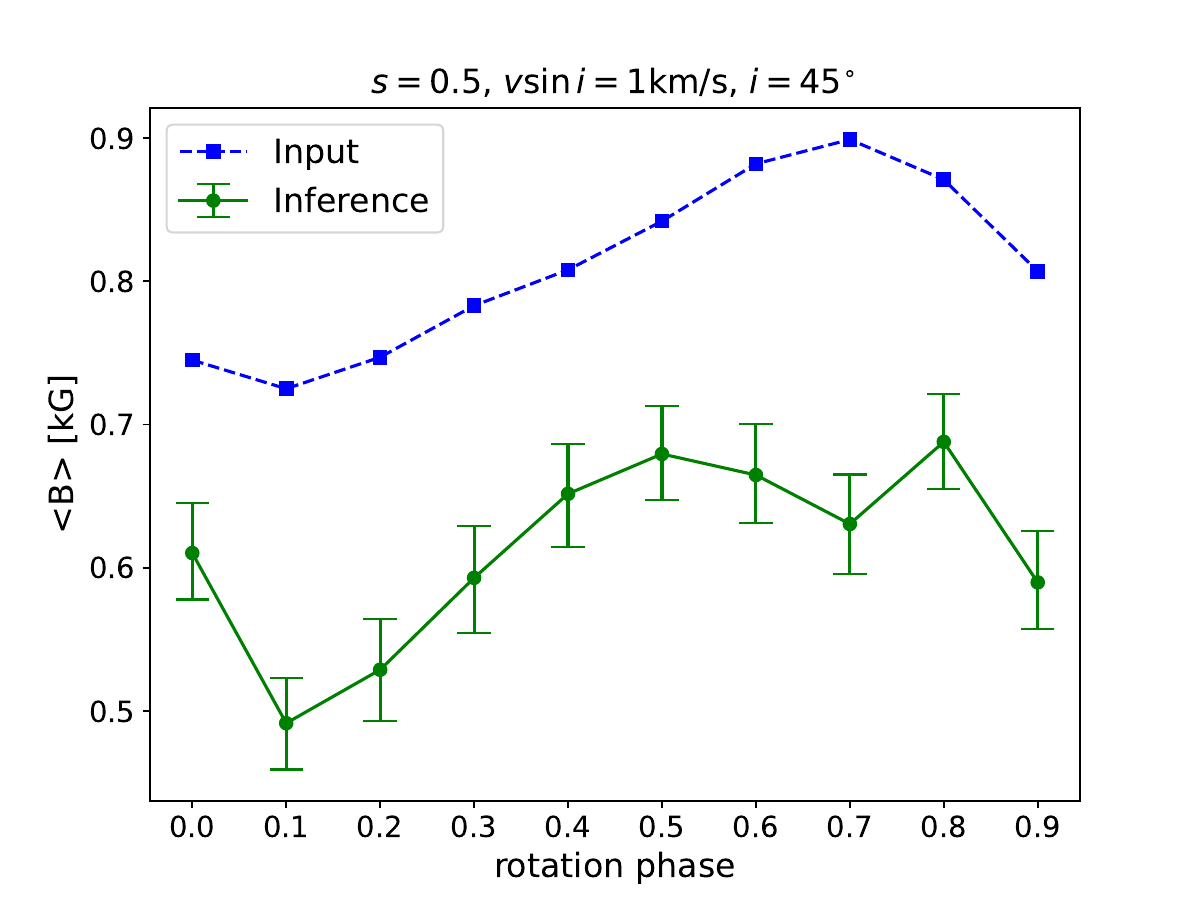}
    \includegraphics[width=0.33\linewidth]{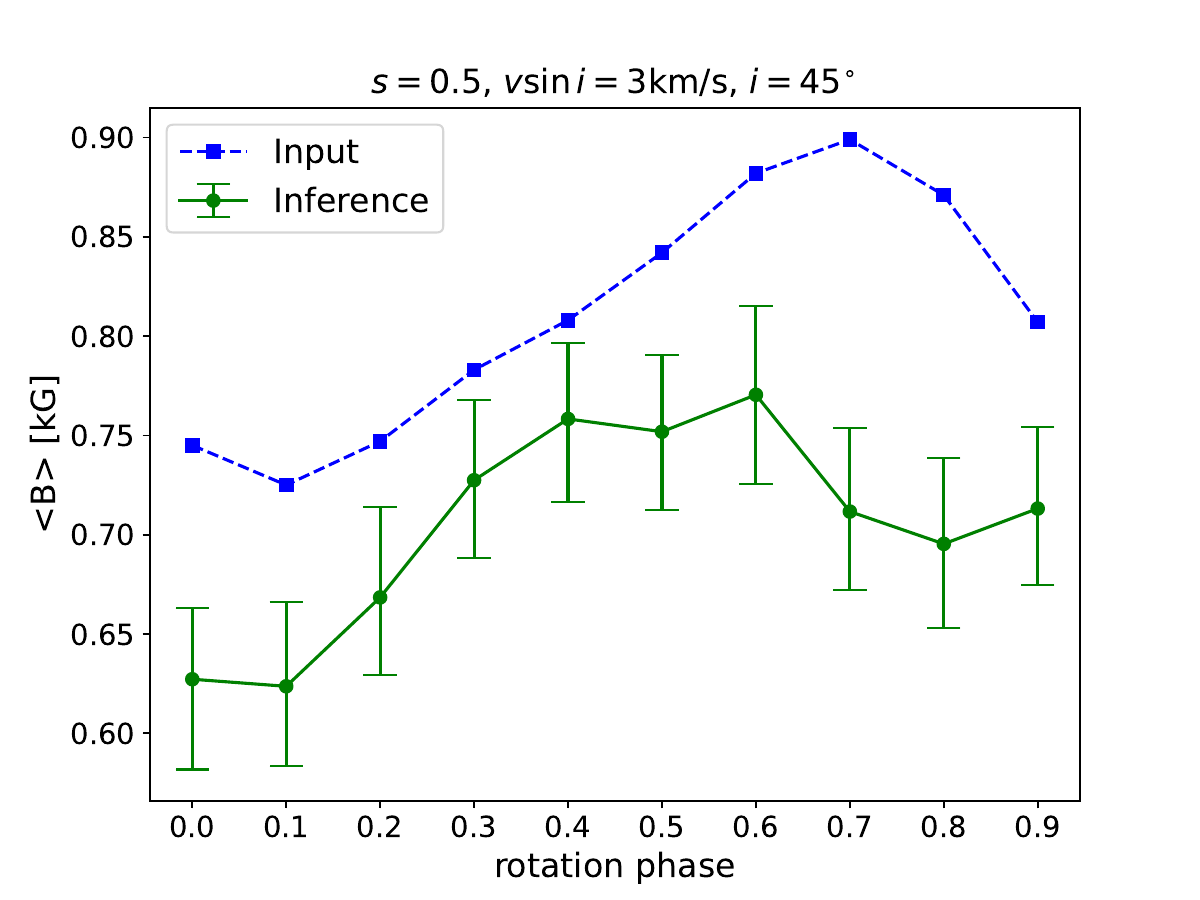}
    \includegraphics[width=0.33\linewidth]{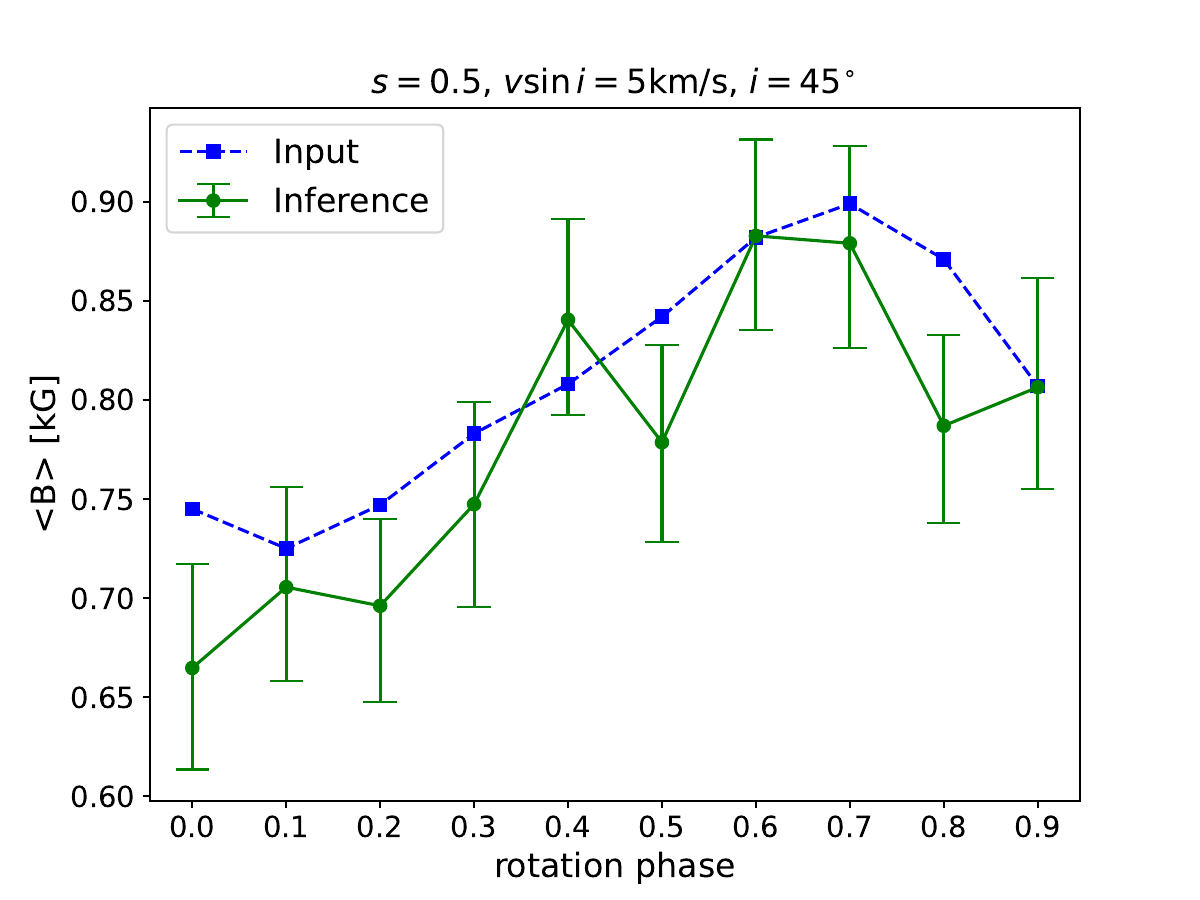}
    \caption{Same as Fig.~\ref{fig:input_recB_s2_v5_i45} for the case of $s=0.5$, \vsini\,=\,1, 3, 5~\kms, $i=45$\degr.}
    \label{fig:input_recB_s05_i45}
\end{figure*}

\begin{figure*}[!h]
    \centering
    \includegraphics[width=0.33\linewidth]{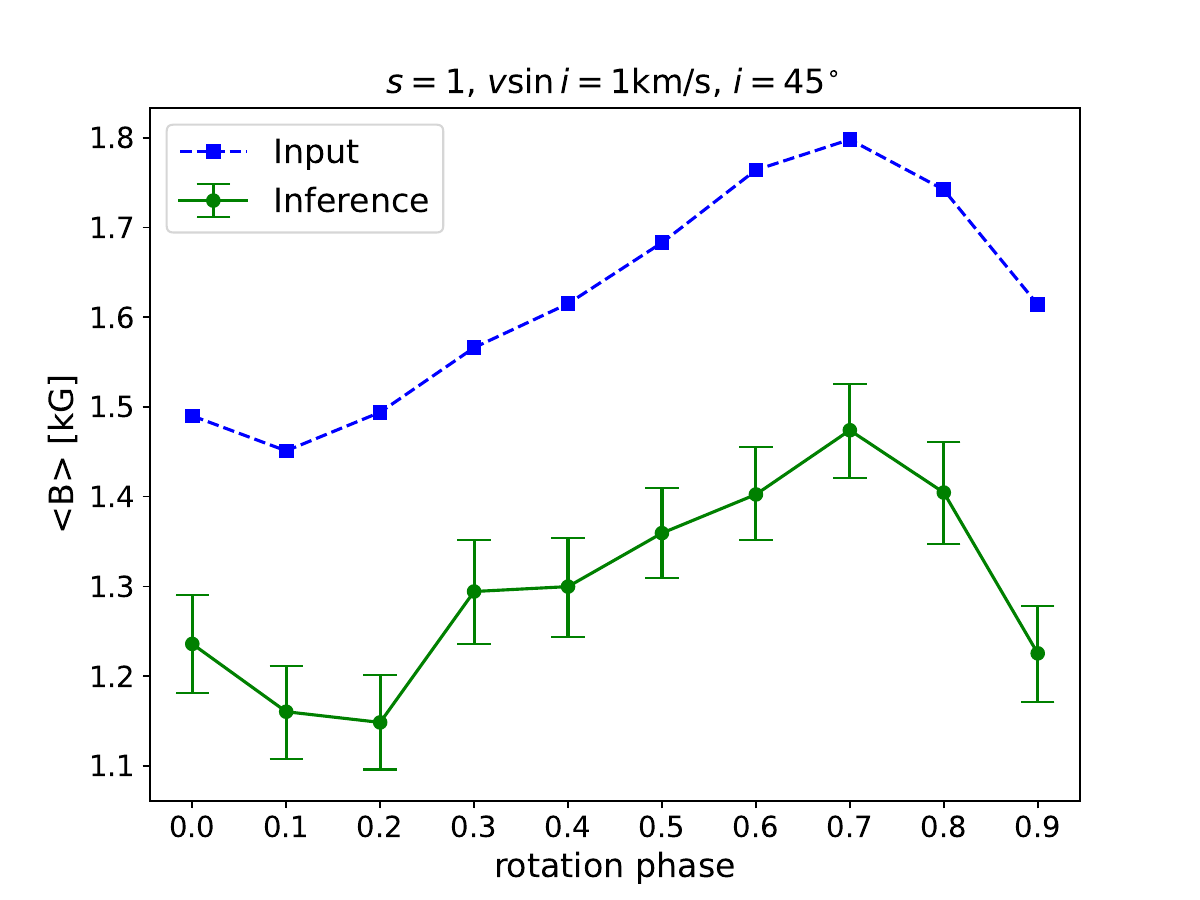}
    \includegraphics[width=0.33\linewidth]{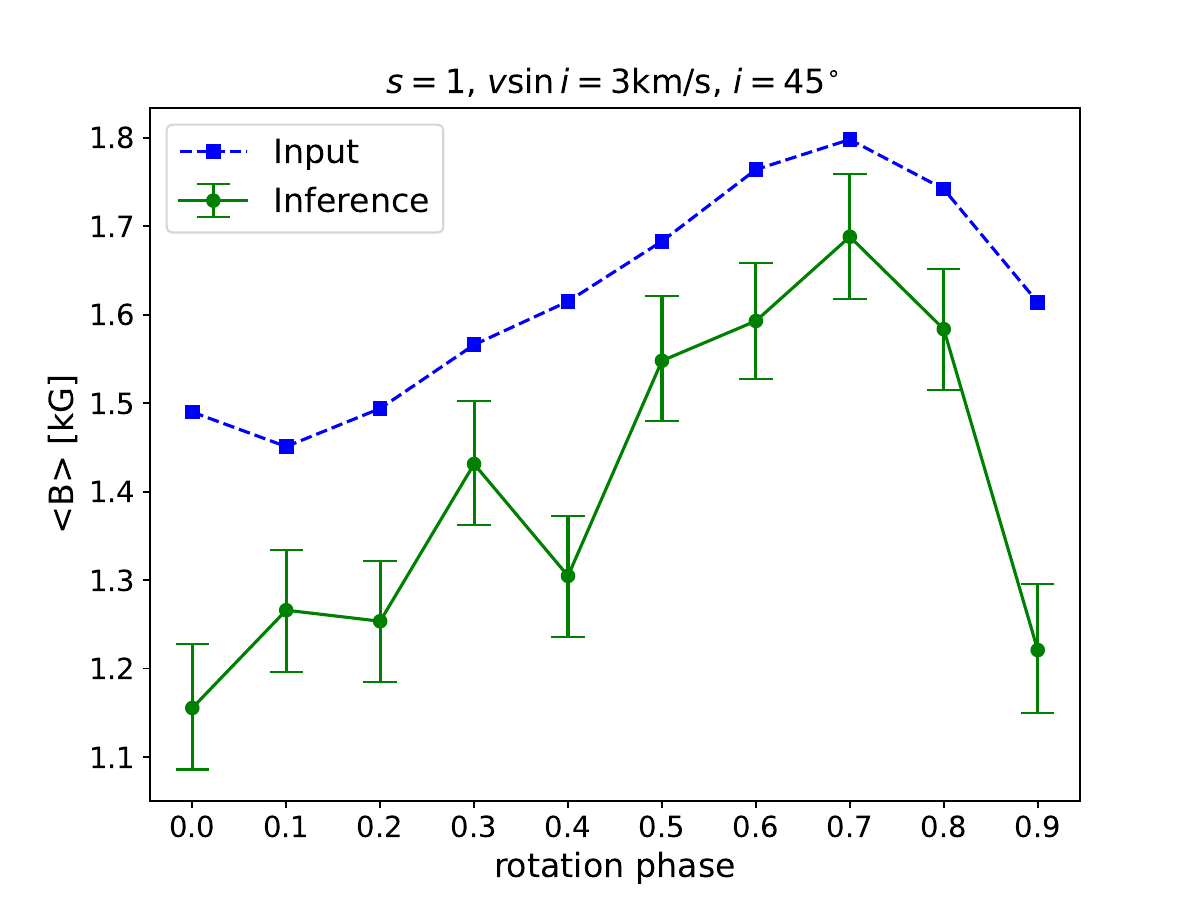}
    \includegraphics[width=0.33\linewidth]{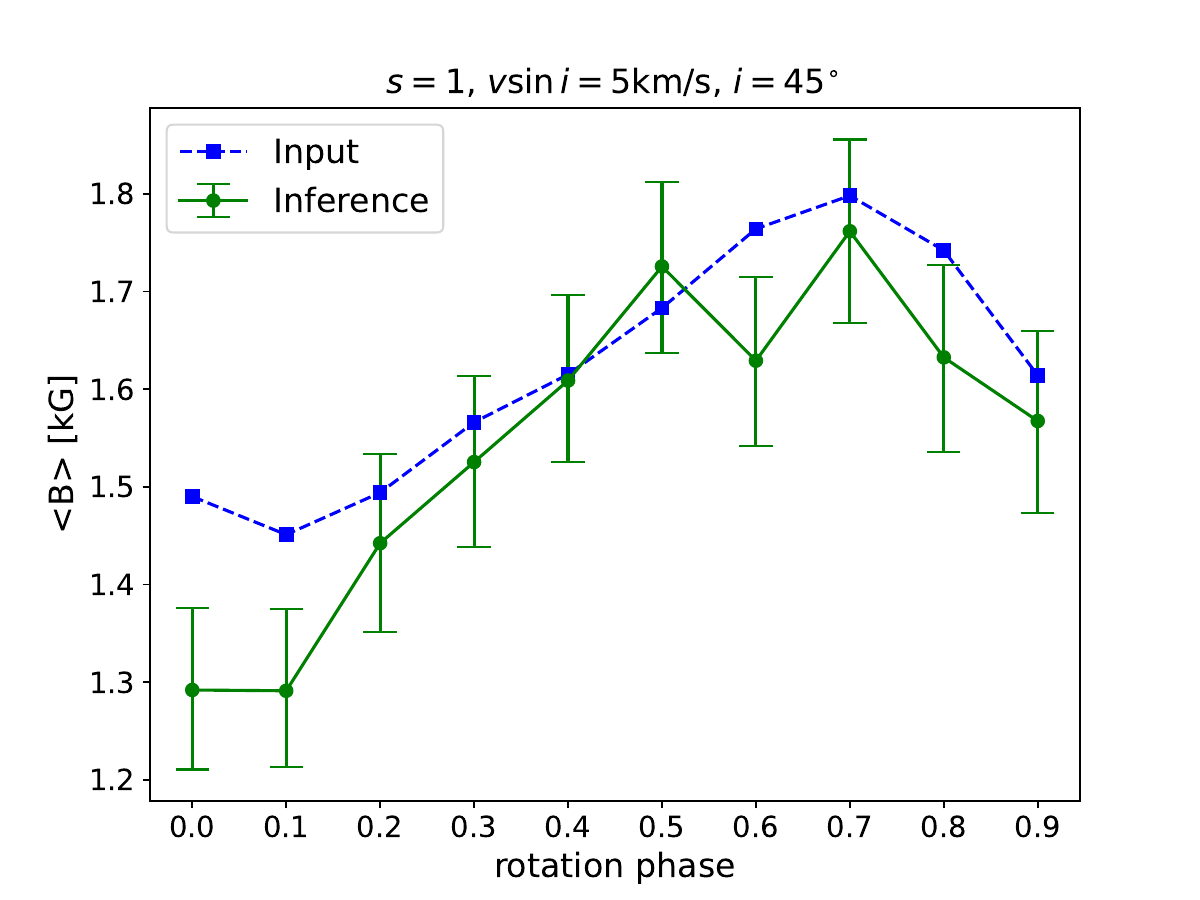}
    \includegraphics[width=0.33\linewidth]{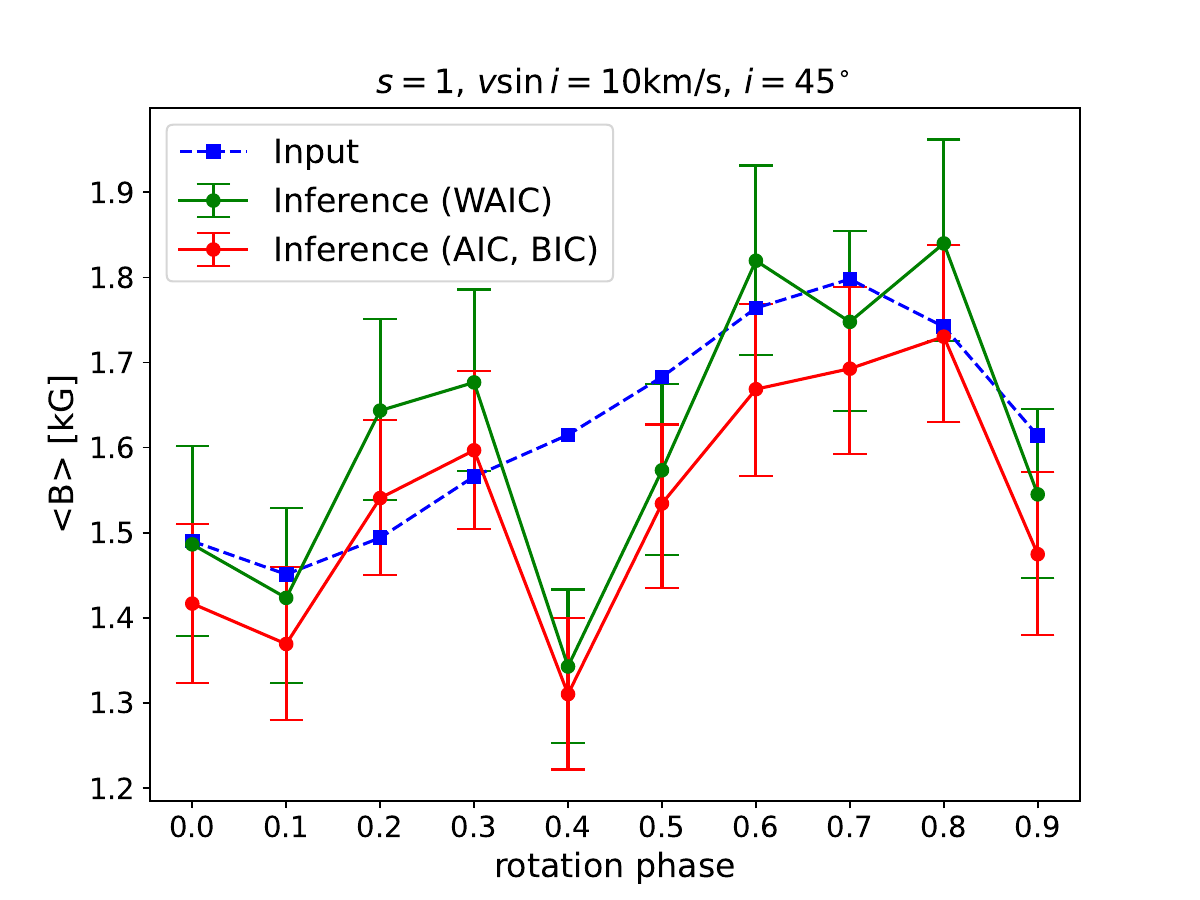}
    \caption{Same as Fig.~\ref{fig:input_recB_s2_v5_i45} for the case of $s=1$, \vsini\,=\,1, 3, 5~\kms, $i=45$\degr.}
    \label{fig:input_recB_s1_i45}
\end{figure*}

\begin{figure*}[!h]
    \centering
    \includegraphics[width=0.33\linewidth]{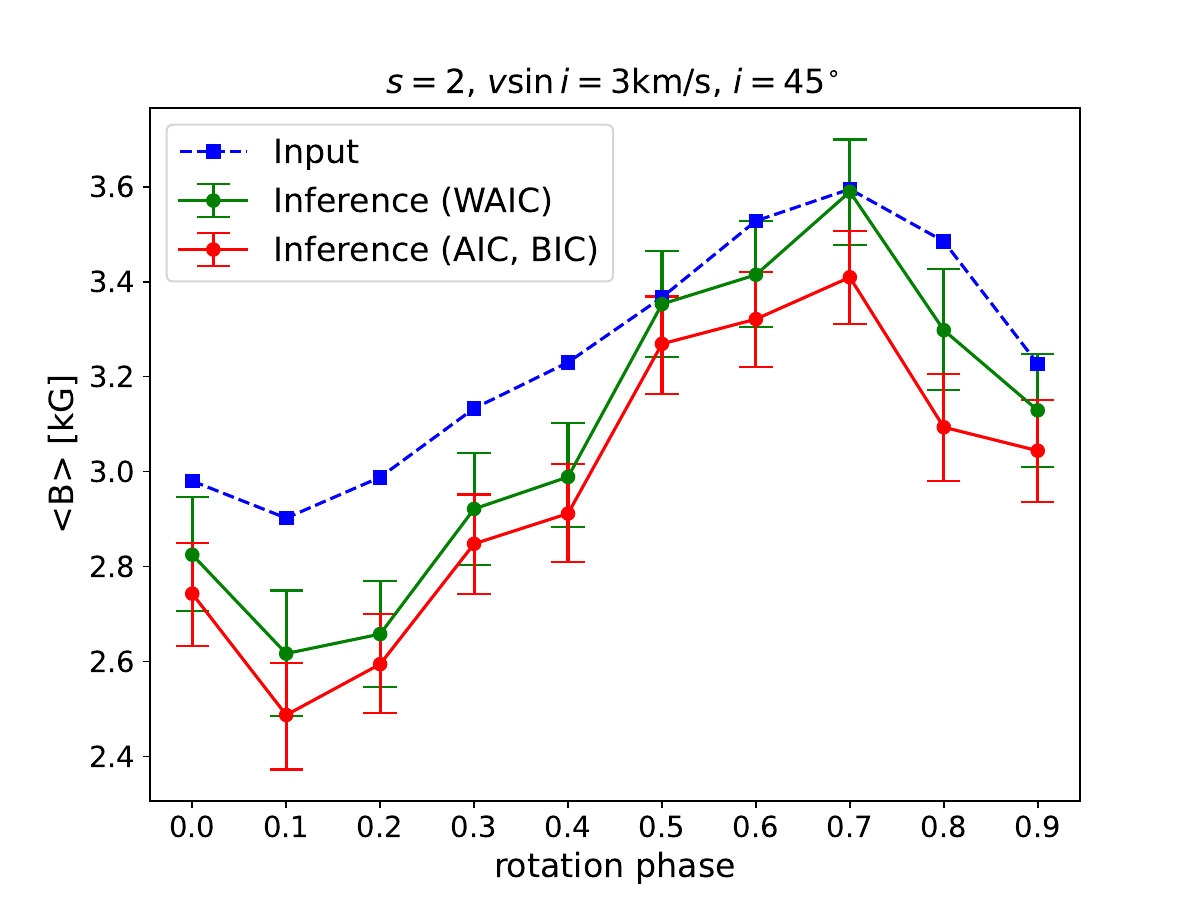}
    \includegraphics[width=0.33\linewidth]{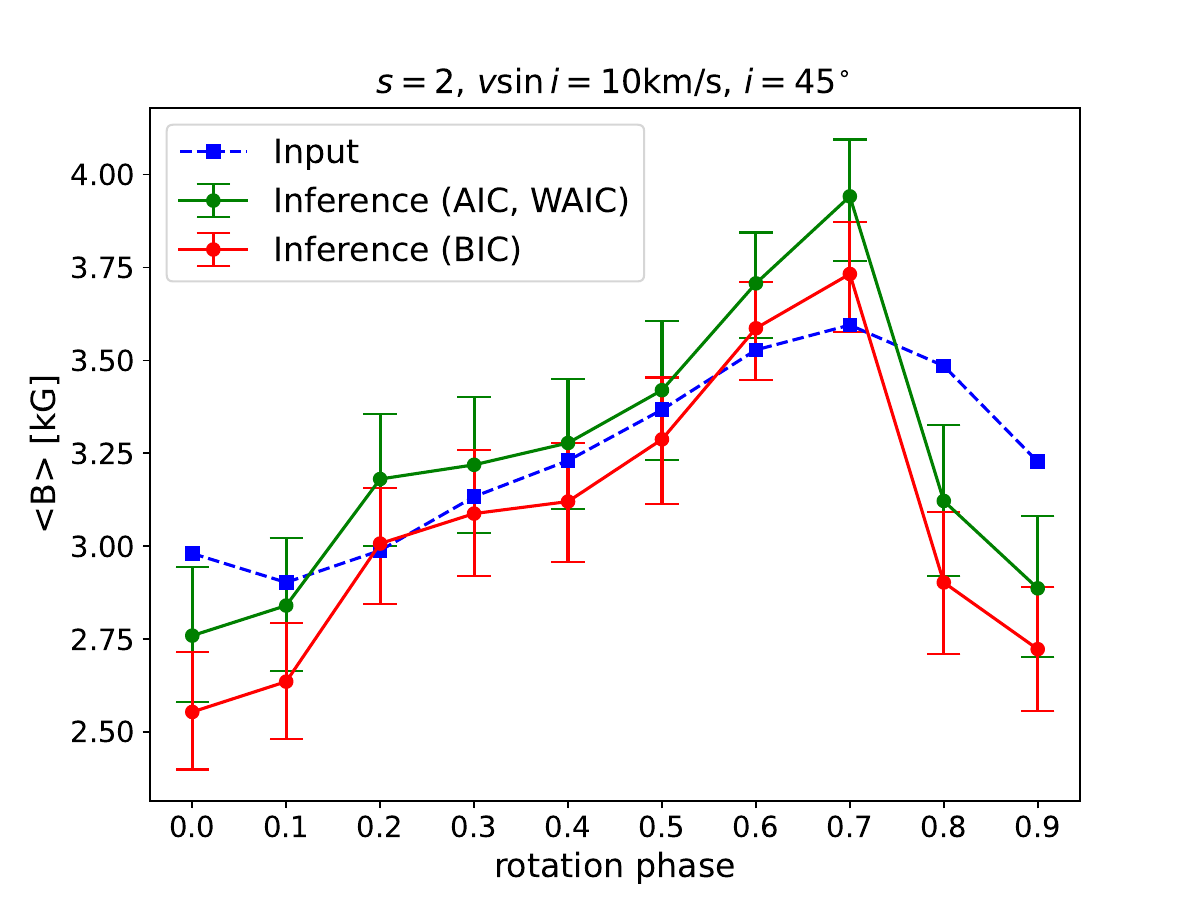}
    \includegraphics[width=0.33\linewidth]{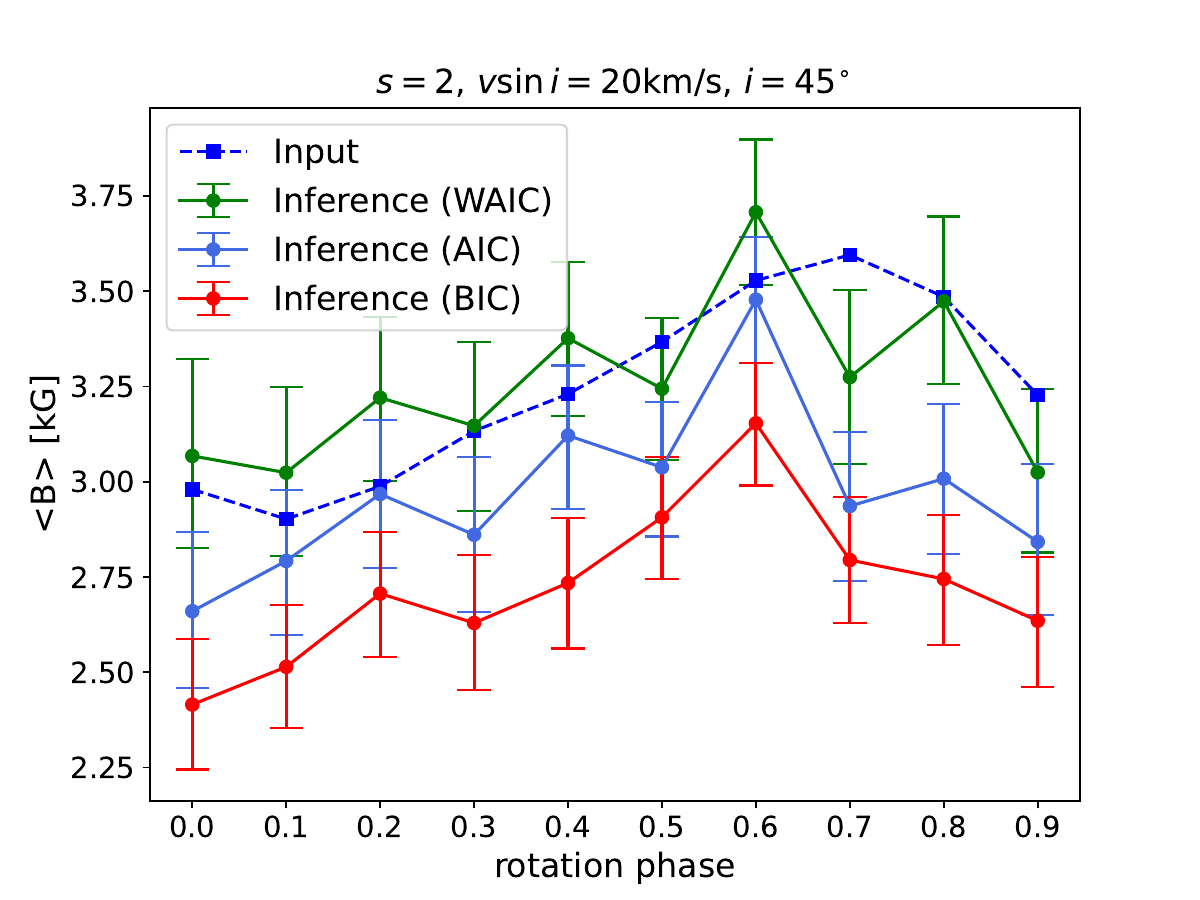}
    \caption{Same as Fig.~\ref{fig:input_recB_s2_v5_i45} for the case of $s=2$, \vsini\,=\,3, 10, 20~\kms, $i=45$\degr.}
    \label{fig:input_recB_s2_i45}
\end{figure*}

\newpage 

\subsection{Inference with different noise realisation in simulated observations for the case of $s=1$, \vsini\,=\, 10~\kms, $i=45$\degr}
\label{appen_different-noise}

The following plots present the results of the inference for the case $s=1$, \vsini\,=\,10~\kms, $i=45$\degr, and a set of simulated observations with a different noise realisation. The inference employed magnetic filling factors defined in steps of 2~kG and different number of magnetic components minimising BIC (red line/markers), AIC (blue), and WAIC (green) (see Sect.~\ref{sub_sec:results_choice_max_field}). The mean hemispheric magnetic field averaged over the rotation phases is \bs$_p=1.55^{+0.10}_{-0.11}$~kG while the mean deviation from the true field strength is $\langle\langle B \rangle - \langle B \rangle_{\mathrm{input}} \rangle_p = -0.07 \pm 0.12$~kG.

\begin{figure*}[!h]
    \centering
    \includegraphics[width=0.33\linewidth]{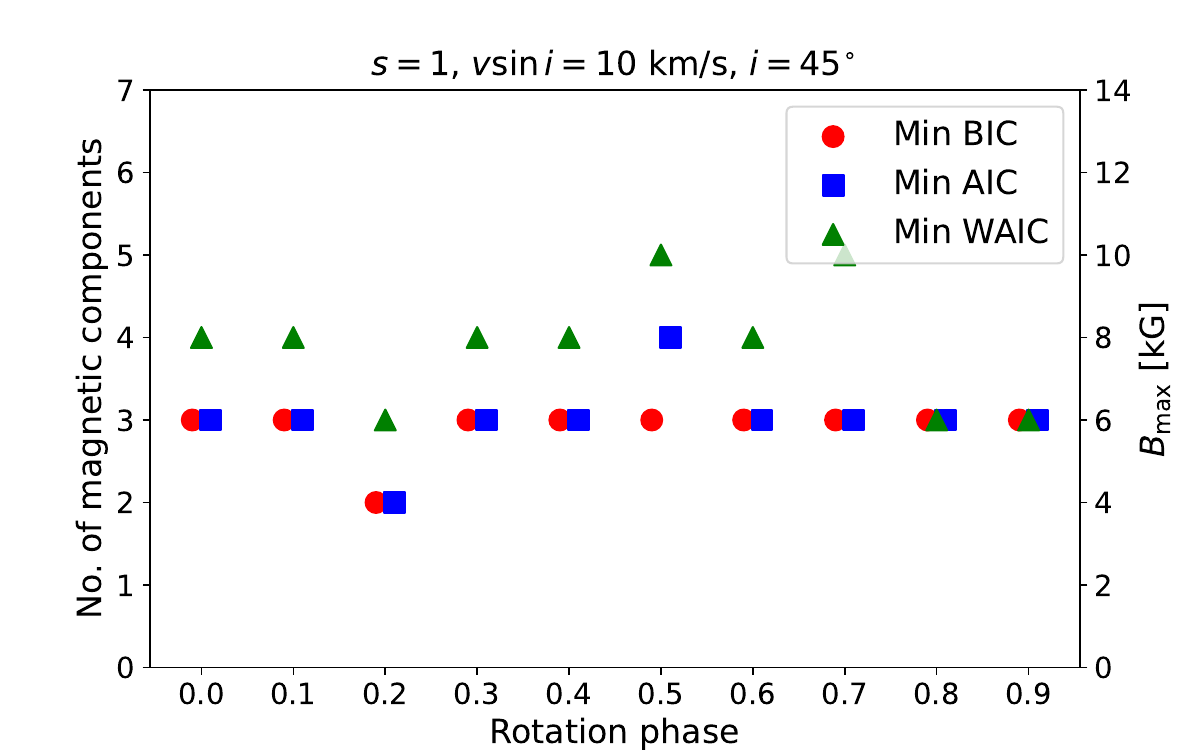}
    \includegraphics[width=0.33\linewidth]{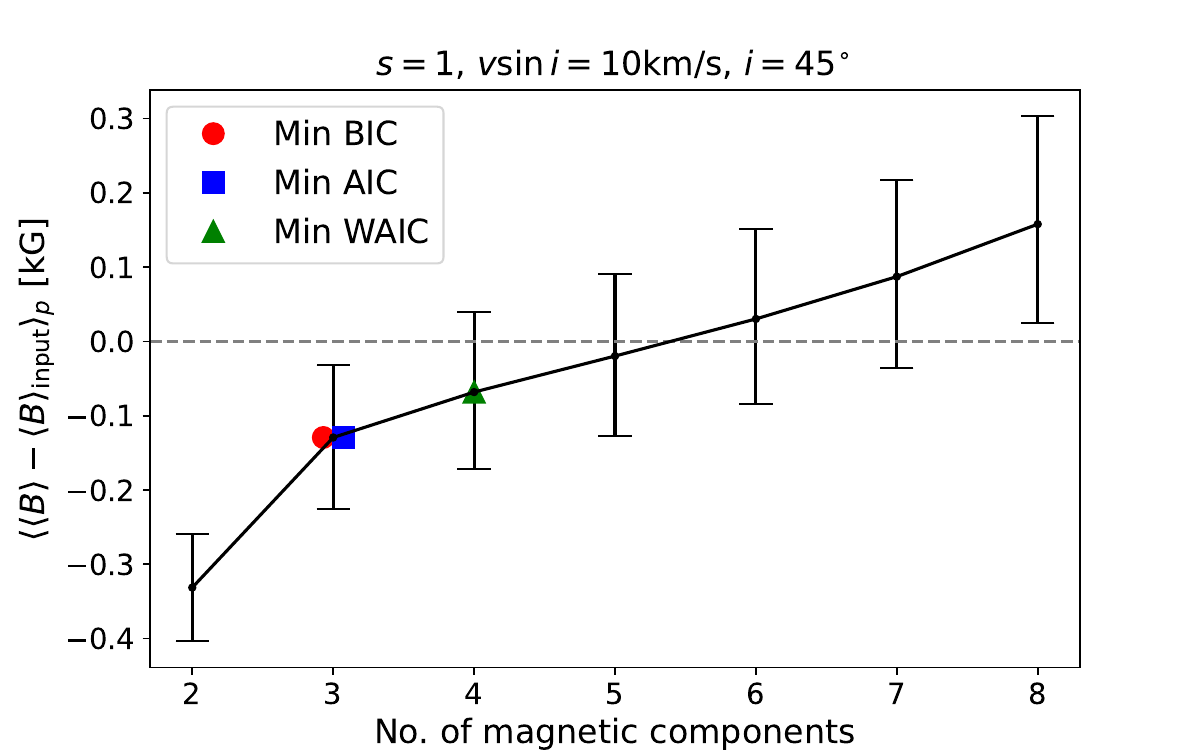}
    \includegraphics[width=0.33\linewidth]{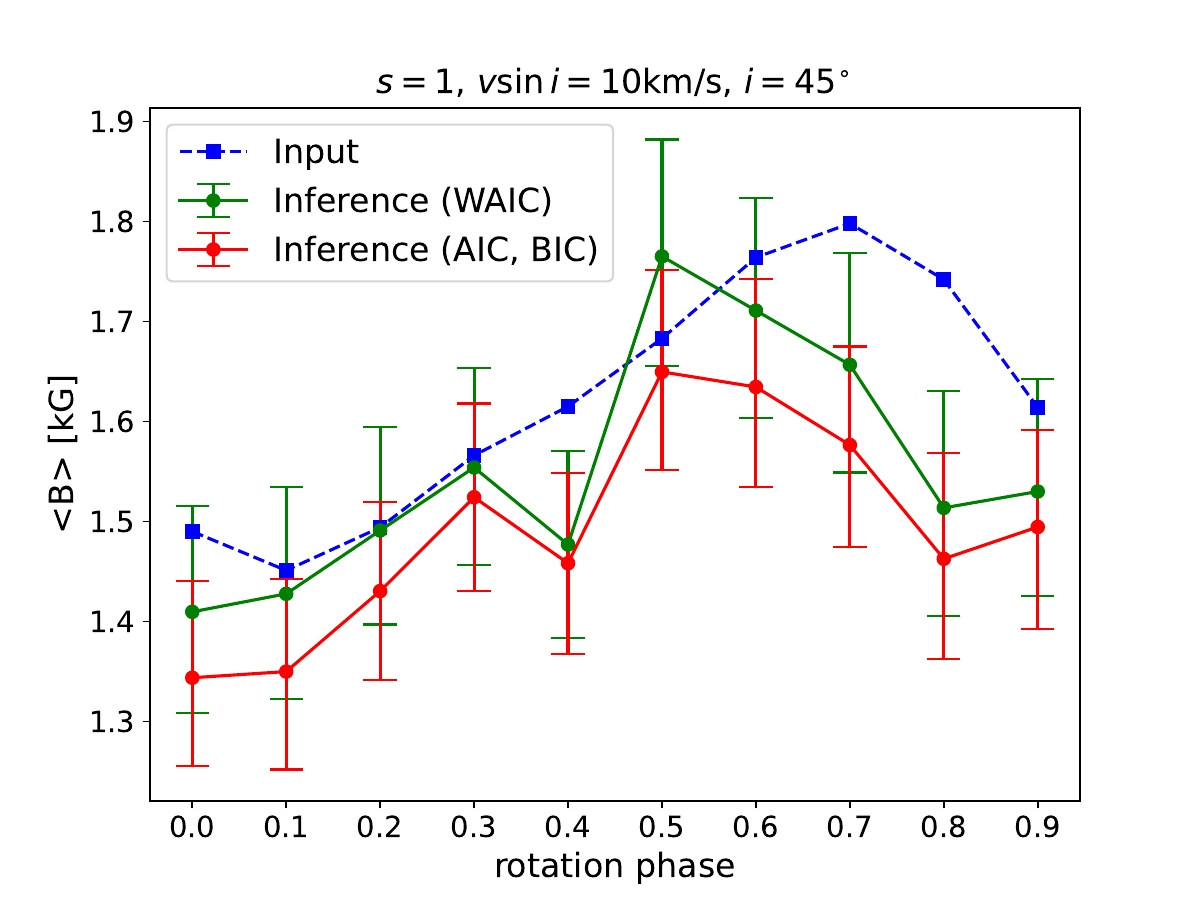}
    \caption{Results of the inference for the case of $s=1$, \vsini\,=\,10~\kms, $i=45$\degr, and simulated observations with a new noise realisation. \textit{Left and centre}: Same as Fig.~\ref{fig:criteria_diff_s1_v10_i45}. \textit{Right}: Same as Fig.~\ref{fig:input_recB_s2_v5_i45}.}
    \label{fig:input_recB_s1_v10_i45_newNoise}
\end{figure*}

\subsection{Inference with alternative fitting strategies and finer field strength grid}

\label{appen_1kG}
The following plots present the inferred values of the hemispheric surface average magnetic field, \bs, with two different line strengths fitting strategies, with comparison to the values extracted from the input magnetic map. The inference employed magnetic filling factors defined in steps of 1~kG and a maximum field strength of 4~kG. See Sect.~\ref{sub_sec:results_comparisonR22} for details of the inference setups.

\begin{figure*}[!h]
    \centering
    \includegraphics[width=0.33\linewidth]{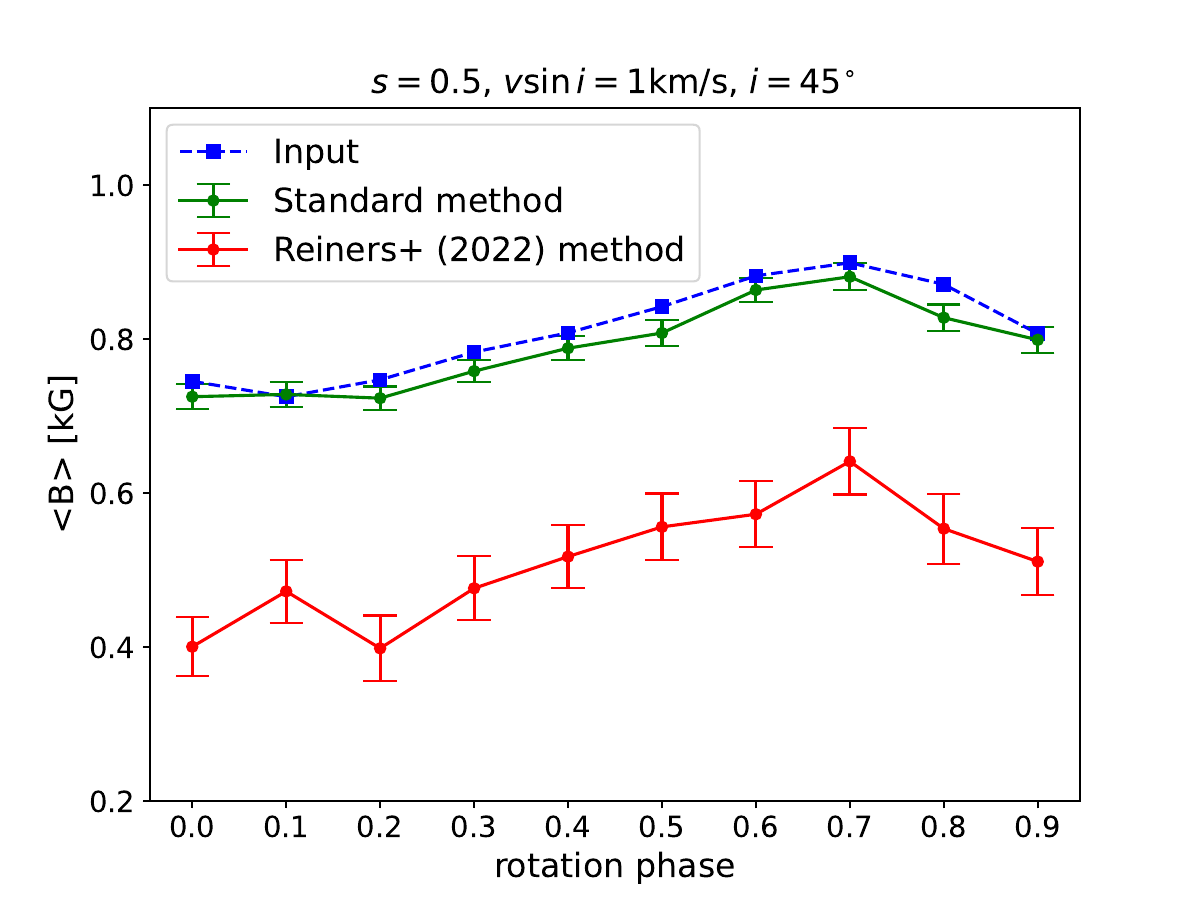}
    \includegraphics[width=0.33\linewidth]{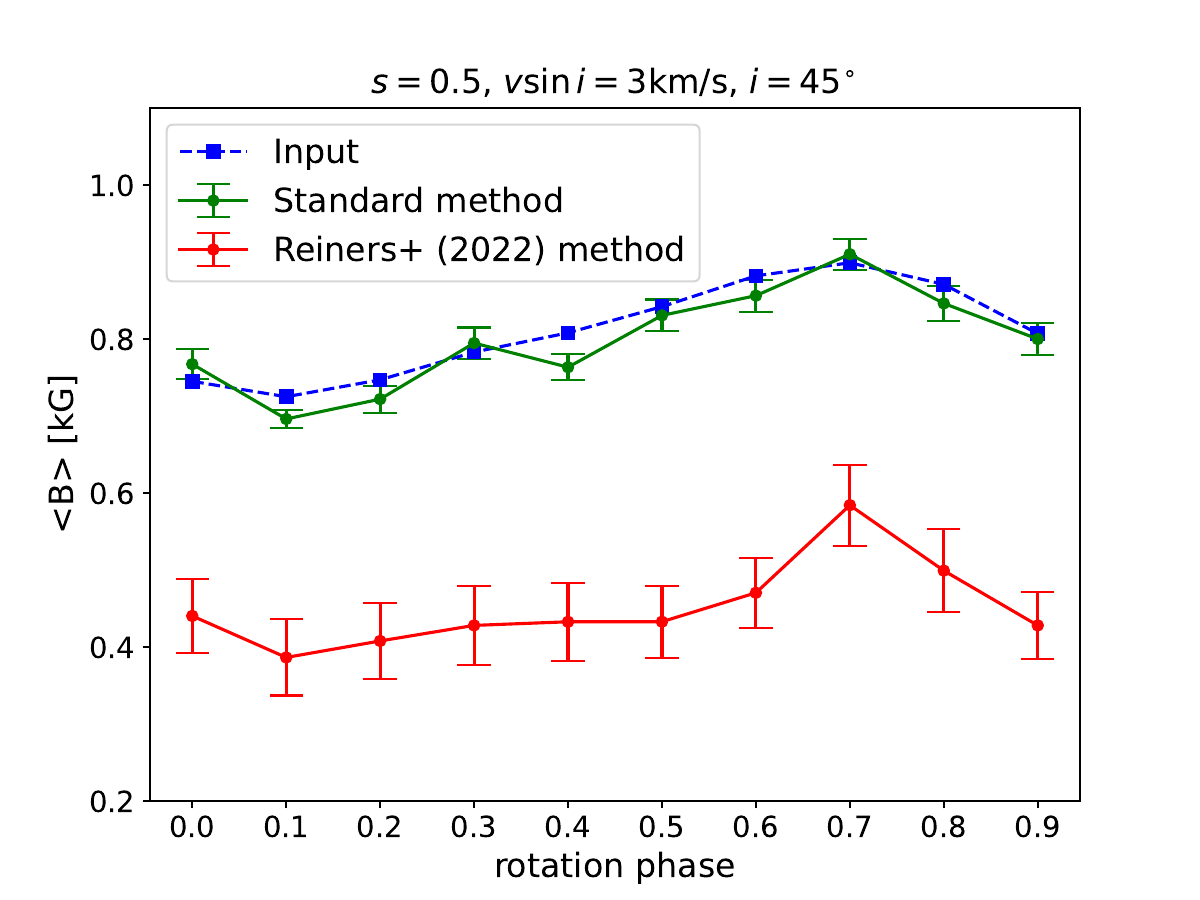}
    \includegraphics[width=0.33\linewidth]{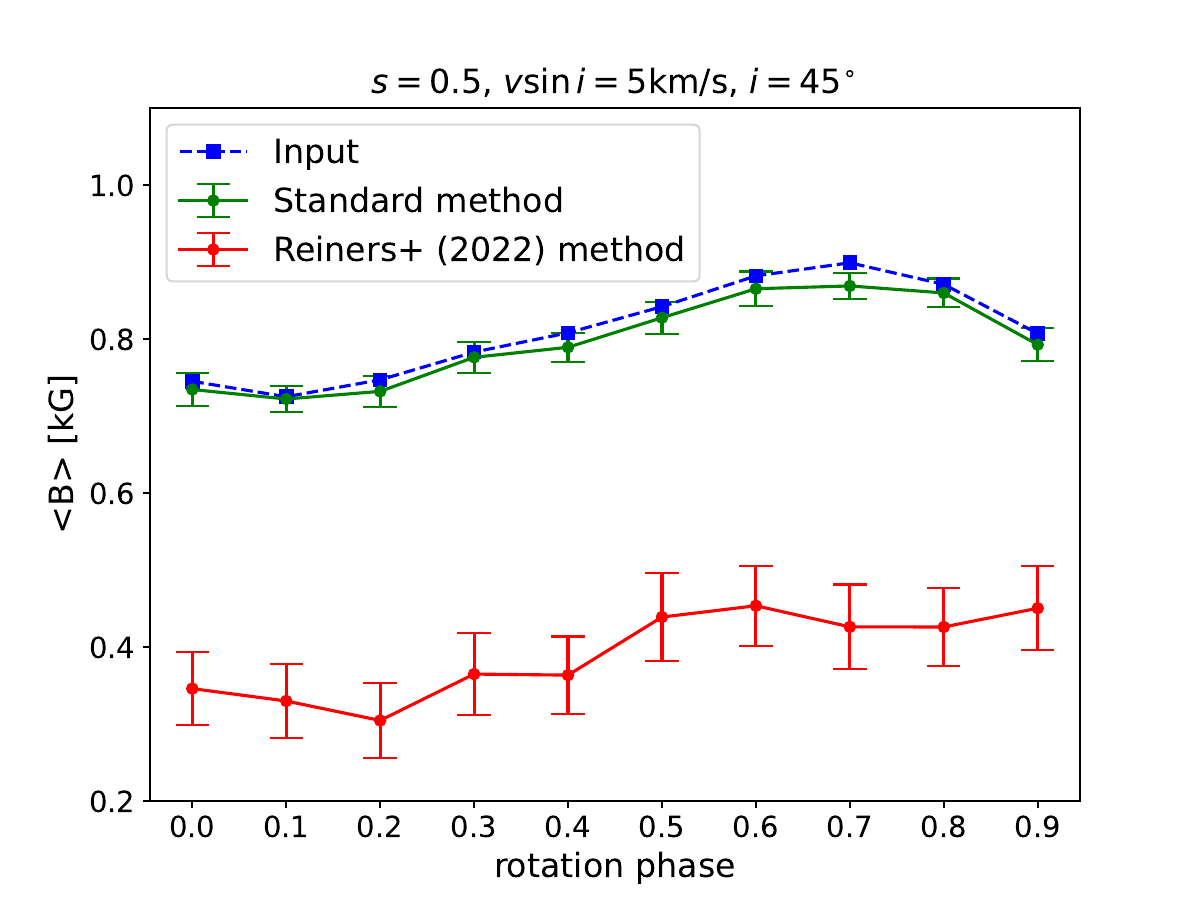}
    \caption{Same as Fig.~\ref{fig:input_recB_s2_v5_i45} for the case of $s=0.5$, \vsini\,=\,1, 3, 5~\kms, $i=45$\degr\ and with the setup of Sect.~\ref{sub_sec:results_comparisonR22}. Green lines are the result of the inference with continuum scaling factors as free parameters and red lines are the result of the inference using \cite{Reiners2022} method with scaling of individual line strengths.}
    \label{fig:input_recB_s05_i45_R22_ourMethod}
\end{figure*}

The following table presents, for the case of $s=0.5$, \vsini\,=\,1~\kms, $i=45$\degr, the inferred rotation-phase averaged mean hemispheric magnetic field and the rotation-phase average of the difference between the inferred and input mean hemispheric fields obtained using different magnetic field discretisations. The inferences employed magnetic filling factors defined with either 1~kG or 2~kG spacing and different numbers of magnetic components. See Sect.~\ref{sub_sec:results_comparisonR22} for details of the inference setups.

\begin{table*}[!h]
\caption{The inferred mean hemispheric field averaged over the rotation phases and corresponding discrepancies with respect to the input field for the case of $s=0.5$, \vsini\,=\,1~\kms, $i=45$\degr, using different step size and number of magnetic components.}  
\label{tab:test-step-size-R22}    
\centering           

\begin{tabular}{C C C C C | c | c}      
\hline\hline            
\multicolumn{5}{c|}{Magnetic components (kG)}& \bs$_p$ & $ \langle\langle B \rangle - \langle B \rangle_{\mathrm{input}} \rangle_p$ \\
0 & 1 & 2 & 3 & 4 & (kG) & (kG)\\
\hline                      
$\bullet$ &   & $\bullet$ &   &   & $0.613 \pm 0.034$ & $-0.198 \pm 0.038$ \\
$\bullet$ &   & $\bullet$ &   & $\bullet$ & $0.618 \pm 0.034$ & $-0.193 \pm 0.038$ \\
$\bullet$ & $\bullet$ & $\bullet$ &   &   & $0.692 \pm 0.010$ & $-0.119 \pm 0.017$ \\
$\bullet$ & $\bullet$ & $\bullet$ & $\bullet$ &   & $0.784 \pm 0.016$ & $-0.027 \pm 0.012$ \\
$\bullet$ & $\bullet$ & $\bullet$ & $\bullet$ & $\bullet$ & $0.790 \pm 0.016$ & $-0.021 \pm 0.012$ \\
\hline                                 
\end{tabular}
\end{table*}

\end{appendix}

\end{document}